\documentclass[final,11pt,superscriptaddress,aps,pre,a4paper]{revtex4}
\pdfoutput=1
\usepackage{latexsym}
\usepackage{times}
\usepackage{graphicx}
\usepackage[tight,FIGBOTCAP]{subfigure}
\usepackage{setspace}

\def\a{\alpha}

\def\d{\delta}

\begin{document}

\title{Mean-field methods in evolutionary duplication-innovation-loss
  models for the genome-level repertoire of protein domains }

\author{A.~Angelini}
\affiliation{Universit\`a degli Studi di Milano, Dip.
    Fisica.  Via Celoria 16, 20133 Milano, Italy} 

\author{A.~Amato}
\affiliation{Universit\`a degli Studi di Milano, Dip.
    Fisica.  Via Celoria 16, 20133 Milano, Italy} 

\author{G.~Bianconi} 
\affiliation{Physics Department, Northeastern University, 111 Forsyth St., 111
Dana Research Center, Boston 02115, MA, USA}

\author{B.~Bassetti}
\affiliation{Universit\`a degli Studi di Milano, Dip.
    Fisica.  Via Celoria 16, 20133 Milano, Italy} 
\author{M.~{Cosentino~Lagomarsino}} 
\affiliation{Universit\`a degli Studi di Milano, Dip.
    Fisica.  Via Celoria 16, 20133 Milano, Italy} 
  \email[ e-mail address: ]{Marco.Cosentino-Lagomarsino@unimi.it}
\altaffiliation{and I.N.F.N. Milano, Italy. Tel. +39 02 50317477 ; fax
  +39 02 50317480 }
\altaffiliation{and I.N.F.N. Milano, Italy. Tel. +39 02 50317477 ; fax
  +39 02 50317480 }

\begin{abstract}
  We present a combined mean-field and simulation approach to
  different models describing the dynamics of classes formed by
  elements that can appear, disappear or copy themselves.
  These models, related to a paradigm duplication-innovation model
  known as Chinese Restaurant Process, are devised to reproduce the
  scaling behavior observed in the genome-wide repertoire of protein
  domains of all known species. In view of these data, we discuss the
  qualitative and quantitative differences of the alternative model
  formulations, focusing in particular on the roles of element loss
  and of the specificity of empirical domain classes.
\end{abstract}

\date{\today}

\maketitle

\section{Introduction}

Understanding the complexity of genomes and the drives that shape them
is a fundamental problem of contemporary biology~\cite{TL04}, which
poses a number of challenges to contemporary statistical mechanics.
Considering this problem from a large-scale viewpoint, the basic
observables to account for are the distributions of the different
``functional components'' (such as genes, introns, non-coding RNA,
etc.)  encoded by sequenced genomes of varying size.

When these genome-wide data are parametrized by measures of ``genome
size'' (such as the number of bases or the number of genes in a genome),
there are important emerging ``scaling laws'' both for the classes of
evolutionary related genes~\cite{HvN98,Mattick2004,Maslov2009}, the
functional categories of genes~\cite{vN03,ITA06} and some non-coding parts
of genomes~\cite{LC03b}.  These scaling laws are the signs of
universal invariants in the processes and constraints that gave rise
to the genomes as they can be observed today. A current challenge is
the understanding of these laws using physical modeling concepts and the
comparison of the models to the available whole-genome data. This effort
can help disentangle neutral from selective effects~\cite{R08,K08}.

Here we consider the statistical features of the set of proteins
expressed by a genome, or proteome. A convenient level of analysis is
a description of the proteome in terms of structural protein
domains~\cite{OT05}. Domains are modular ``topologies'', or
sub-shapes, forming proteins~\cite{BT99}. A domain determines a set
of potential biochemical or biophysical functions and interactions for
a protein, such as binding to other proteins or DNA and participation
in well-defined classes of biochemical reactions~\cite{OT05,ITA06}.
Despite the practically unlimited number of possible protein
sequences, the repertoire of basic topologies for domains seems to be
relatively small~\cite{KWK02}.
With a loose parallel, domains could be seen as an ``alphabet'' of
basic elements of the protein universe.  Understanding the usage of
domains across organisms is as important and challenging as decoding
an unknown language.

The content of a genome is determined primarily by its evolutionary
history, in which neutral processes and natural selection play
interdependent roles.  In particular, the coding parts of genomes
evolve by some well-defined basic ``moves'': gene loss, gene
duplication, horizontal gene transfer (the transfer of genetic
material between unrelated species), and gene genesis (the \emph{de
  novo} origin of genes).  Since domains are modular evolutionary
building blocks for proteins, they are coupled to the dynamics
followed by genes. In particular, a new domain topology can emerge by
genesis or horizontal transfer, and new domains of existing domain
topologies can emerge by duplication or be lost.  Finally, topologies
can be completely lost by a genome if the last domain that carries
them is lost (see Fig.~\ref{fig:scheme}).

Large-scale data concerning structural domains are available from
bioinformatic databases, and can be analyzed at the genome level.
These coarse-grained data structures can be represented as sets of
``domain classes'' (the sets of all realizations of the same domain topology
in proteins), populated by domain realizations. In particular, much
attention has been drawn by the intriguing discovery that the
population of domain classes have power-law
distributions~\cite{QLG01,KWR+02,Kuz02,AD05,DSS02}: the number
$F(j,n)$ of domain classes having $j$ members follows the power-law
$\sim 1/ j^{z}$, where the exponent $z$ typically lies between 1 and
2.
An interesting thread of modeling work ascribes the emergence of
power-laws to a generic preferential-attachment principle due to gene
duplication. Growth models are formulated as nonstationary,
duplication-innovation models~\cite{QLG01,KLQ+06,DS05} and as
stationary birth-death-innovation
models~\cite{KWR+02,KWK03,KWB+04,KBK05}.

Intriguingly, as we have recently shown~\cite{Lagomarsino2009}, the domain
content of genomes also exhibits
scaling laws as a function of the total number of domains $n$,
indicating that even evolutionarily distant genomes show common trends
where the relevant parameter is their size.
\begin{itemize}
\item[(i)] The number of domain classes (or distinct hits of the same
  domain) concentrates around a master curve $F(n)$ that appears
  to be markedly sublinear with size, perhaps saturating.
\item[(ii)] The fitted exponent of the power-law-like distribution
  $F(j,n)$ of domain classes having $j$ members, in a proteome of size
  $n$ decreases with genome size. In other
  words, there is evidence for a cutoff that increases linearly with
  $n$.
\item[(iii)] The occurrence of fold topologies across genomes is
  highly inhomogeneous - some domain superfamilies are found in
  all genomes, some rare, with a sigmoid-like drop between these two
  categories~\footnote{See ref.~\cite{Lagomarsino2009}, these trends are also
visible from Fig.~\ref{fig:agreement}, ~\ref{fig:alphafit}, ~\ref{fig:cutoff},
and ~\ref{fig:occurrence} of this work.}.
\end{itemize}

We recently reported the above collective trends, and showed how the
scaling laws in the data could be reproduced using universal
parameters with non-stationary duplication-innovation models.
Our results indicate that the basic evolutionary
moves themselves can determine the observed scaling behavior of
domain content, \emph{a priori} of more specific biological trends.

This modeling approach, while similar in formulation to that of
previous investigators~\cite{QLG01} who did not consider these scaling
laws, has important modifications, mostly related to the scaling with
$n$ of the relative probability of adding a domain belonging to a new
class and duplicating an existing one. To reproduce the observed
trends, newly added domain classes cannot be treated as
\emph{independent} random variables, but are conditioned by the
preexisting proteome structure.

In this paper, we give a detailed account of this modeling approach,
considering different variants of duplication-innovation-loss models
for protein domains, and relate them to available results in the
mathematical and physical literature.  In particular, we will focus on
mean-field approaches for the models and comparison with direct
simulation, and we will show how they can be generally used to obtain
the main qualitative and quantitative trends.

The first part of the paper is devoted to the minimal model
formulation, which only includes duplication and innovation moves for
domains, and relates to the so-called \emph{Chinese Restaurant
  Process (CRP)} of the mathematical literature~\cite{Pit02,PY97}.  We
will review the main known results for this model, derive analytically
solvable mean-field equations, and show how they compare to the
available rigorous results and the finite size behavior.

The rest of the paper is devoted to biologically motivated variants of
the main model related to two main features: including the role of
loss of domains, which is a frequently reported event, and breaking
the exchange symmetry of domain classes, which is unrealistic, as
specific protein domains perform different biological functions.  For
these variants, we will present mean-field and simulations results,
and characterize their phenomenology in relation with empirical
data.  In particular, while in general for these variants the rigorous
mathematical results existing for the CRP break down, we will show
how the use of simple mean-field methods proves to be a robust tool
for accessing the qualitative phenomenology.

\newpage

\section{Generic features of the model}

\begin{figure}[htb]
  \centering
  \includegraphics[width=0.8\textwidth]{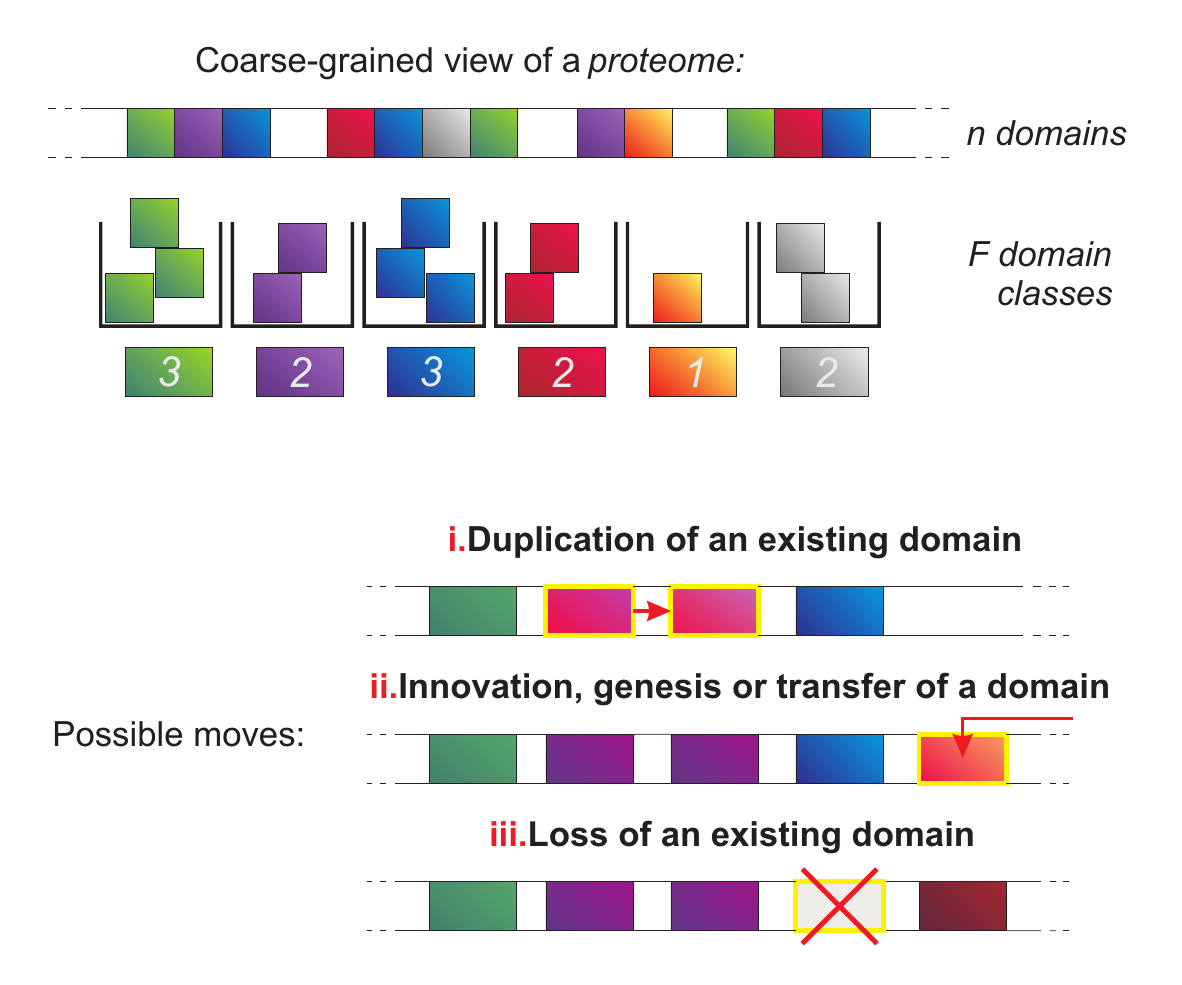}
  \caption{(Color online) Scheme of the generic features of the
duplication-innovation-loss model. Top: The sequences of proteins coded by a
    genome can be broken down into domains, represented by colored
    boxes. All the boxes of the same color, representing domains with
    the same topology are collected in the same domain class. Bottom: basic
moves. Elements of a class can be
    duplicated or lost, and new classes can be formed with prescribed
    probabilities.}
  \label{fig:scheme}
\end{figure}

The model represents a proteome through its repertoire of domains.
Domains having the same topology are collected in domain classes
(Fig.~\ref{fig:scheme}). Thus the relevant data structures are
partitions of elements (domains) into classes.
The basic observables considered  are the following: $n$,
the total number of domains, $f(n)$, a random variable indicating the
number of classes (distinct domain topologies) at size $n$, a random
variable $k_i$, the population of class $i$, $n_i$, the size at birth
of class $i$, and $f(j,n)$: the number of domain classes having $j$
members. We will generally indicate mean values by capitalized
letters (e.g.\ $F(n)$ is the mean value of $f(n)$, $K_i$ the mean
value of $k_i$, etc.).
 
The model is conceived as a stochastic process based on the elementary
moves available to a genome (Fig.~\ref{fig:scheme}) of adding and
losing domains, associated to relative probabilities: $p_O$, the
probability to duplicate an old domain (modeling gene duplication),
$p_N$, the probability to add a new domain class with one member
(which describes domain innovation, for example by horizontal
transfer), and $p_L$, a loss probability (which we will initially
disregard, and consider in a second step).
Iteratively, either a domain is added or it is lost with the prescribed
probabilities.

An important feature of the duplication move is the (null) hypothesis
that duplication of a domain has uniform probability along the genome,
and thus it is more probable to pick a domain of a larger class. This
is a common feature with previous models~\cite{QLG01}. This hypothesis
creates a ``preferential attachment''~\cite{BA99} principle, stating
the fact that duplication is more likely in a larger domain class,
which, in this model as in previous ones, is responsible for the
emergence of power-law distributions.

In mathematical terms, if the duplication probability is split as the
sum of per-class probabilities $p_O^i$, this hypothesis
requires that $p_O^i \propto k_i$, where $k_i$ is the population of
class $i$, i.e.\ the probability of finding a domain of a particular
class and duplicating it is proportional to the number of members of
that class.

It is important to notice that in this model, while $n$ can be used as
an arbitrary measure of time, the ratio of the time-scales of
duplication and innovation is not arbitrary, and is set by the ratio
$p_N/p_O$. In the model of Gerstein and coworkers, this is taken as a
constant, as the innovation move considered to be statistically
independent from the genome content. In particular, both probabilities
are considered to be constant. This choice has two problems.  First,
it cannot give the observed sublinear scaling of $F(n)$.  Indeed, if
the probability of adding a new domain is constant with $n$, so will
be the rate of addition, implying that this quantity will increase on
average linearly with genome size.  Moreover, the same model
gives power-law distributions for the classes with exponent larger than
two, in contrast with most of the available data.

Previous investigators did not consider the fact that genomes cluster
around a common curve~\cite{Lagomarsino2009},
and thought of each of them as coming from an independent stochastic
process with different parameters~\cite{QLG01}.  Furthermore, choosing
constant $p_N$ implies that for larger genomes the influx of new
domain families is heavily dominant on the flux of duplicated domains.

As noted by Durrett and Schweinsberg~\cite{DS05}, constant $p_N$ makes
sense only if one thinks that new fold topologies emerge from an
internal ``nucleation-like'' process with constant rate, rather than
from an external flux. This process could be pictured as the genesis
of new topologies from sequence mutation.  Empirically, while genesis
events are reported~\cite{Kunin2003} and must occur, it is clear that
domain topologies are very stable, and the exploration of sequence space is not
free, but conditioned by a number of additional important factors,
including chromosomal position and expression patterns of genes, and
their role in biological networks~\cite{PPL06}.
Moreover, in prokaryotes, it is known that a large contribution to the
innovation of
coding genomes is provided by horizontal gene
transfer~\cite{Kunin2003}, the exchange of genetic material between
species, which can be reasonably represented in a model as an
external flux, as opposed to the internal nucleation process
representing genesis.

For eukaryotes, horizontal gene transfer is less important, and there
can be multiple relevant innovation processes including exonization,
loss of exons, alternative start sites changing the protein. We have
not attempted to model the detailed processes leading to innovation,
and because of their higher complexity in eukaryotes, we prefer to
compare the model to the prokaryote data set alone.  However, we can
point out that in principle the same model has good agreement with the
set of prokaryotes and eukaryotes together~\cite{Lagomarsino2009}.  In
eukaryotes, the change in number of classes with respect to size
change is generally small, but seems to have a trend that "glues"
quite well with prokaryotes (and in particular, innovation decreases
with size).

Motivated by the sublinear scaling of the number of domain classes,
and taking into account in an effective way the role of processes that
condition the addition of new domain topologies, we consider
statistically \emph{dependent} moves.
On general grounds, if a genome is a complex system where
sub-components interact in clusters and non-locally, domain topologies
as well have to be coordinated with other parts of the system, so that
it is reasonable that evolutionary moves are conditioned by what is
already present, and that the actual number of domain topologies need
not to be trivially an extensive quantity.

The simplest way to implement this choice is to concentrate on the
innovation process.  Let us consider the indicator $\xi(n)$, taking
value $1$ if a new domain class is born at size $n$, and value $0$
otherwise.  The number of classes at size $n$ will be $f(n)= \sum_{j
  \leq n} \xi(j)$.  If the random variables $\xi(n)$ are independent
and identically distributed, i.e.\ $p(\xi(n)=1)=p_N =$constant, $f(n)$
follows a Bernoulli distribution whose mean value is linear in $n$.
Moreover, $\frac{f(n)}{n}$ will be increasingly concentrated with
increasing $n$ on the deterministic value $p_N$.  If vice versa the
random variables $\xi(n)$ are statistically dependent or also
simply not identically distributed, the mean value may not be linear
in $n$, and the concentration phenomenon may not occur.  Both features,
dependence and lack of concentration,
are important. The former is necessary to obtain the observed
sublinear behavior, the latter might create an intrinsic ``diversity''
in the genome ensemble, independently on the finite size of observed
genomes (however, the currently available data are insufficient to
establish this empirically).

\section{Simplest formulation: Chinese Restaurant Process}

We investigate this process using analytical asymptotic equations and
simulations.  We start by considering only growth moves, by
duplication and innovation, postponing the inclusion of domain loss in
the model.  We will see that the resulting model contains the basic
qualitative phenomenology of the scaling laws and can thus be regarded
as the paradigmatic case.
One can arrive at the defining equations with different arguments. A
simple way is to assume that domain duplication is a rare event,
described by a Poisson distribution with characteristic time $\tau$,
during which there is a flux of external or new domain topologies
$\frac{\theta }{\tau}$. Then $p_N= \frac{\theta}{n+\theta}$. In this
case the variables $\xi(n)$ are independent but not identically
distributed.  It is immediate to verify that $f(n)$ has mean value
given by $ \sum_{j=1}^n \frac{\theta}{\theta+j}$ and thus grows as
$\sim \theta \log n$. The same result can be obtained by thinking of
domain addition as a dependent move, conditioned on $n$, $f(n)$ or
both.

It is possible to consider different intermediate scenarios where the
pool of old domain classes is in competition with the universe
explorable by the new classes.  The simplest scheme, which turns out
to be quite general, can be obtained by choosing the conditional
probability that a new class is born $(\xi(n+1)=1)$ given the fact
that  $f(n)=f$ at size $n$
\begin{equation}
p_N  = \frac{\theta + \alpha f}{n +\theta} \ \ ,
\end{equation}
hence
\begin{equation}
p_O = \frac{n -  \alpha f}{n +\theta} \ \ ,
\end{equation}
where $\theta \geq 0$ and $\alpha \in [0,1]$.

Considering per-class duplication probability, one can choose
the following expression, that asymptotically establishes the
preferential attachment principle:

\begin{equation}
p_O^i = \frac{k_i-\alpha}{n + \theta}\ \ .
\end{equation}
Here, $\theta$ represents a characteristic number of
domain classes needed for the preferential attachment principle to set
in, and defines the behavior of $f(n)$ for small $n$ ( $n \rightarrow
0$).  $\alpha$ is the most important parameter, which sets the scaling
of the duplication/innovation ratio. 
Intuitively, the smaller $\alpha$, the more the growth of $f$ is
depressed with growing $n$, and since $p_N$ is asymptotically
proportional to the class density $f/n$ it is harder to add a new
domain class in a larger, or more heavily populated genome.  As we
will see, this implies $p_N/p_O \rightarrow 0$ as $n \rightarrow
\infty$, corresponding to an increasingly subdominant influx of new
fold classes at larger sizes. This choice reproduces the sublinear
behavior for the number of classes and the other scaling laws
described in properties (i-iii).

This kind of model has previously been explored in statistics under
the name of Pitman-Yor, or Chinese Restaurant Process
(CRP)~\cite{Pit02,PY97,Ald85,Kin75}, where it is known as one of the
paradigmatic processes that generate partitions of elements into
classes that are symmetric by swapping, or ``exchangeable''. This
process is used in Bayesian inference and clustering
problems~\cite{Qin2006}.  In the Chinese restaurant (with table sharing)
parallel,
individual domains correspond to customers and tables are domain
classes.  A domain belonging to a given class is a customer sitting at
the corresponding table.  In a duplication event, a new customer is
seated at a table with a preferential attachment principle, and in an
innovation event, a new table is added.

\section*{Mean-field theory for the CRP}
A simple mean-field treatment of the CRP allows to access its scaling
behavior.
Rigorous results for the probability distribution of the fold usage
vector $\{k_1,...,k_F\}$, for $f(n)=F$, are in good agreement with
mean-field predictions.  It is important to note that for this
stochastic process, the usual large-deviation theorems do not hold, so
that large-$n$ limit values of quantities such as $\frac{f(n)}{F(n)}$
do not converge to numbers, but rather to random
variables~\cite{Pit02}.
Despite of this non-self-average property, it is possible to
understand the scaling of the averages $K_i$ and $F$ (of $k_i$ and $f$
respectively) at large $n$, writing simple ``mean-field'' equations,
for continuous $n$.  Note that rigorously the mean value $ K_i(n) $ is
still a random variable, function of the (stochastic) birth time $n_i$
of class $i$.

From the definition of the model, we obtain
\begin{equation}
 \partial_n K_i(n) =
\frac{K_i(n)-\alpha}{n+\theta}
\end{equation}
\begin{equation}
\partial_n F(n) = \frac{\alpha
  F(n)+\theta}{n+\theta} \ \ .
\end{equation}
These equations have to be solved with
initial conditions $K_{i}(n_i) = 1$,
and $F(1)=1$. Hence, for $\alpha \ne 0$, one
has
\begin{equation}
 K_i(n) = (1-\alpha) \frac{n+\theta}{n_i+\theta} +\theta \ \ ,
\end{equation}
and
\begin{equation}
  F(n) = \frac{1}{\alpha}
  \left[(\alpha+\theta)\left(\frac{n+\theta}{\theta}\right)^{\alpha}
    -\theta \right] \sim n^{\alpha} \ \ ,
\end{equation}
while, for $\alpha = 0$,
\begin{equation}
  F(n)= \theta \log (n+ \theta) \sim \log(n) \ \ .
\end{equation}
These results imply that the expected asymptotic scaling of $F(n)$ is
sublinear, in agreement with observation (i).

The mean-field solution can be used to compute the asymptotic of
$P(j,n) = F(j,n)/F(n)$, following the same line of reasoning used by
Barabasi and Albert for the preferential attachment model~\cite{BA99}.
This works as follows. From the solution, $j> K_i(n)$ implies $n_i >
n^*$, with $n^*=\frac{(1-\alpha)n -\theta (j-1)}{j-\alpha}$, so that
the cumulative distribution can be estimated by the ratio of the
(average) number of domain classes born before size $n^*$ and the
number of classes born before size $n$, $P(K_i(n) > j) =
F(n^*)/F(n)$. $P(j,n)$ can be obtained by derivation of this
function. For $n,j \rightarrow \infty$, and $j/n$ small, we find
\begin{equation}
  P(j,n)    \sim j^{-(1+\alpha)} \ \ ,
\end{equation}
for $\alpha \ne  0$, and
\begin{equation}
 P(j,n) \sim  \frac{\theta}{j} \ \ ,
\end{equation}
for $\alpha = 0$. The above formulas indicate that the average
asymptotic behavior of the distribution of domain class populations is
a power law with exponent between $1$ and $2$, in agreement with
observation (ii).  In contrast, the behavior of the model of Gerstein
and coworkers~\cite{QLG01,KLQ+06,DS05} can be found in this framework
by taking improperly $\alpha=1$, that is for constant $p_N,\ p_O$. It
gives a linearly increasing $F(n)$ and a power-law distribution with
asymptotic exponent $2$ for the domain classes.
Note that the phenomenology of the Barabasi-Albert preferential
attachment scheme~\cite{BA99} is reproduced by a CRP-like model where
at each step a new domain class (corresponding to the new network
node) with on average $m$ members (the edges of the node) is
introduced, and at the same time $m$ domains are duplicated (the edges
connecting old nodes to the newly introduced one).

\begin{figure}[htbp]
   \centering
   \subfigure[]{
        \label{sub:f1n_emp}
        \includegraphics[width=.4\textwidth]{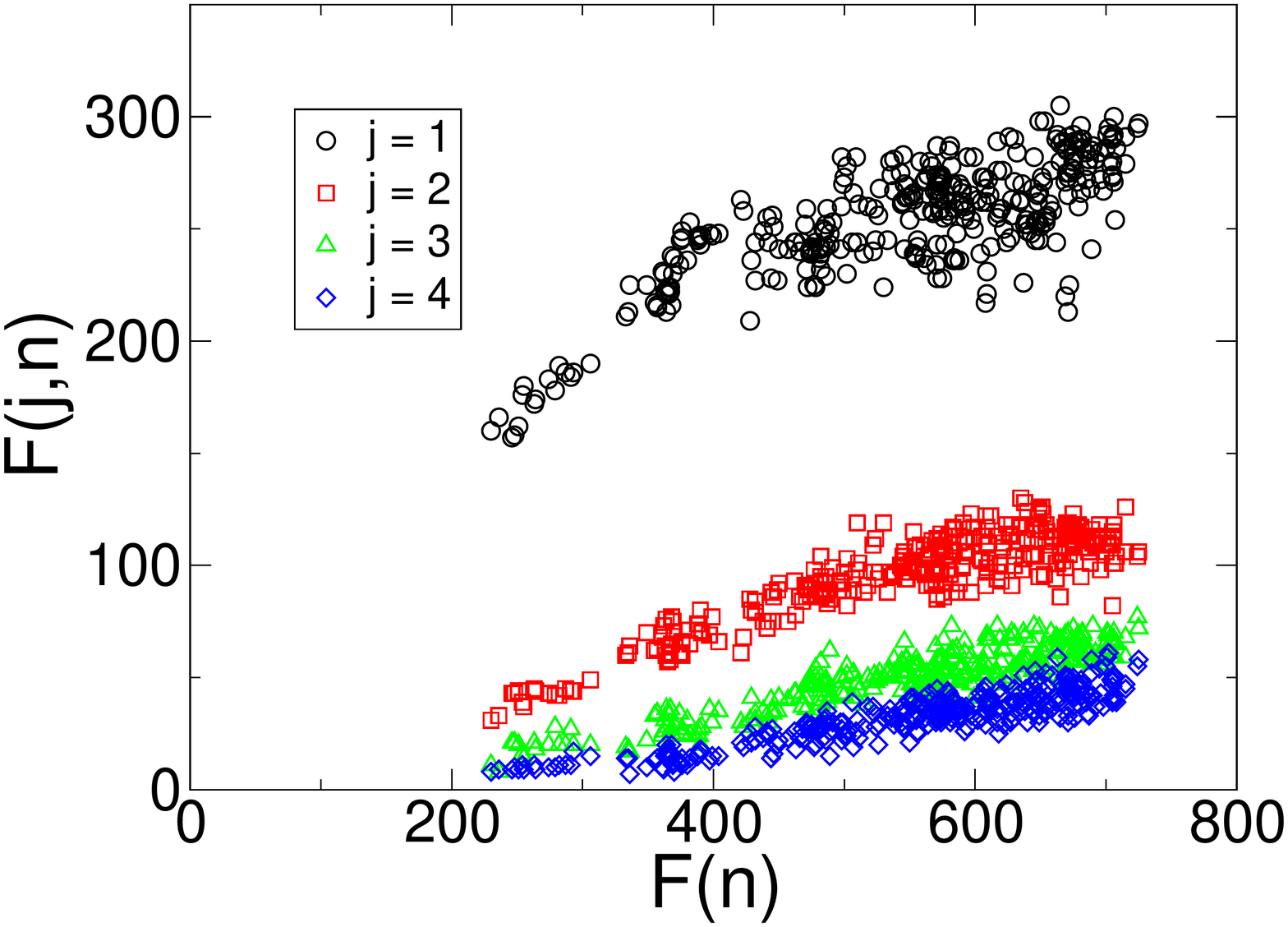}}
   \subfigure[]{
        \label{sub:f1n_d0}
        \includegraphics[width=.4\textwidth]{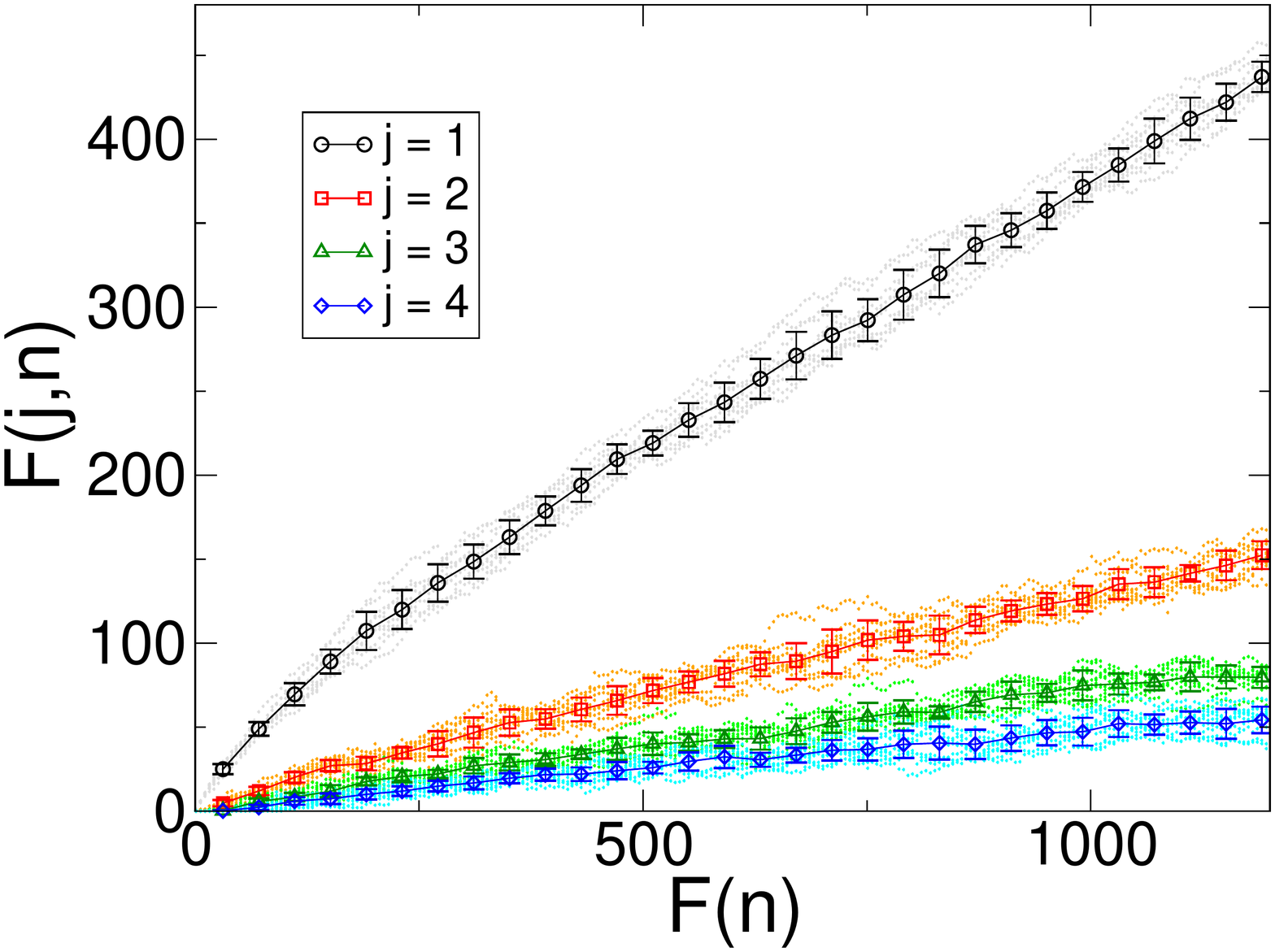}}
   \subfigure[]{
        \label{sub:f1n_m1}
        \includegraphics[width=.4\textwidth]{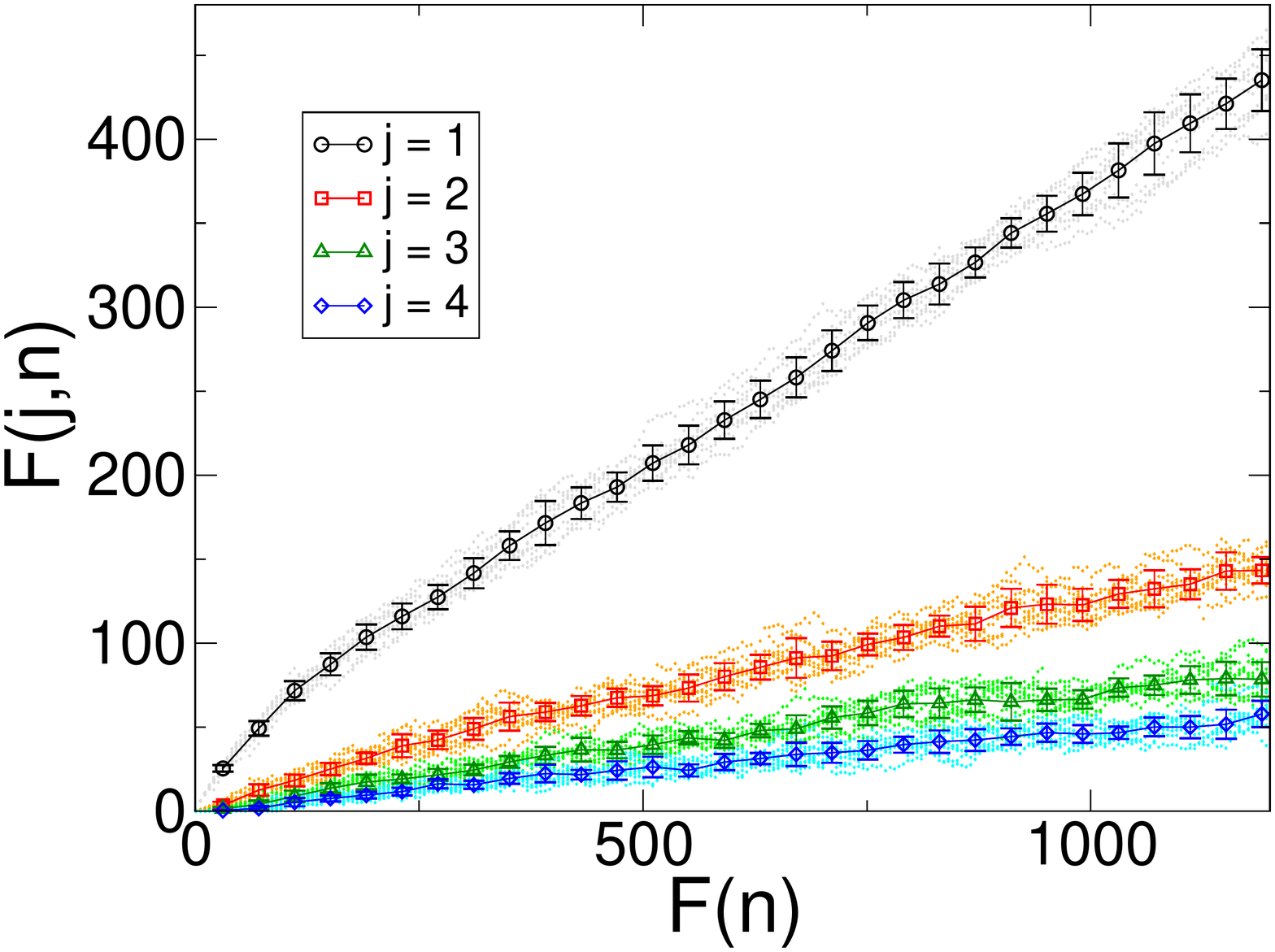}}
  \subfigure[]{
        \label{sub:f1n_m2}
        \includegraphics[width=.4\textwidth]{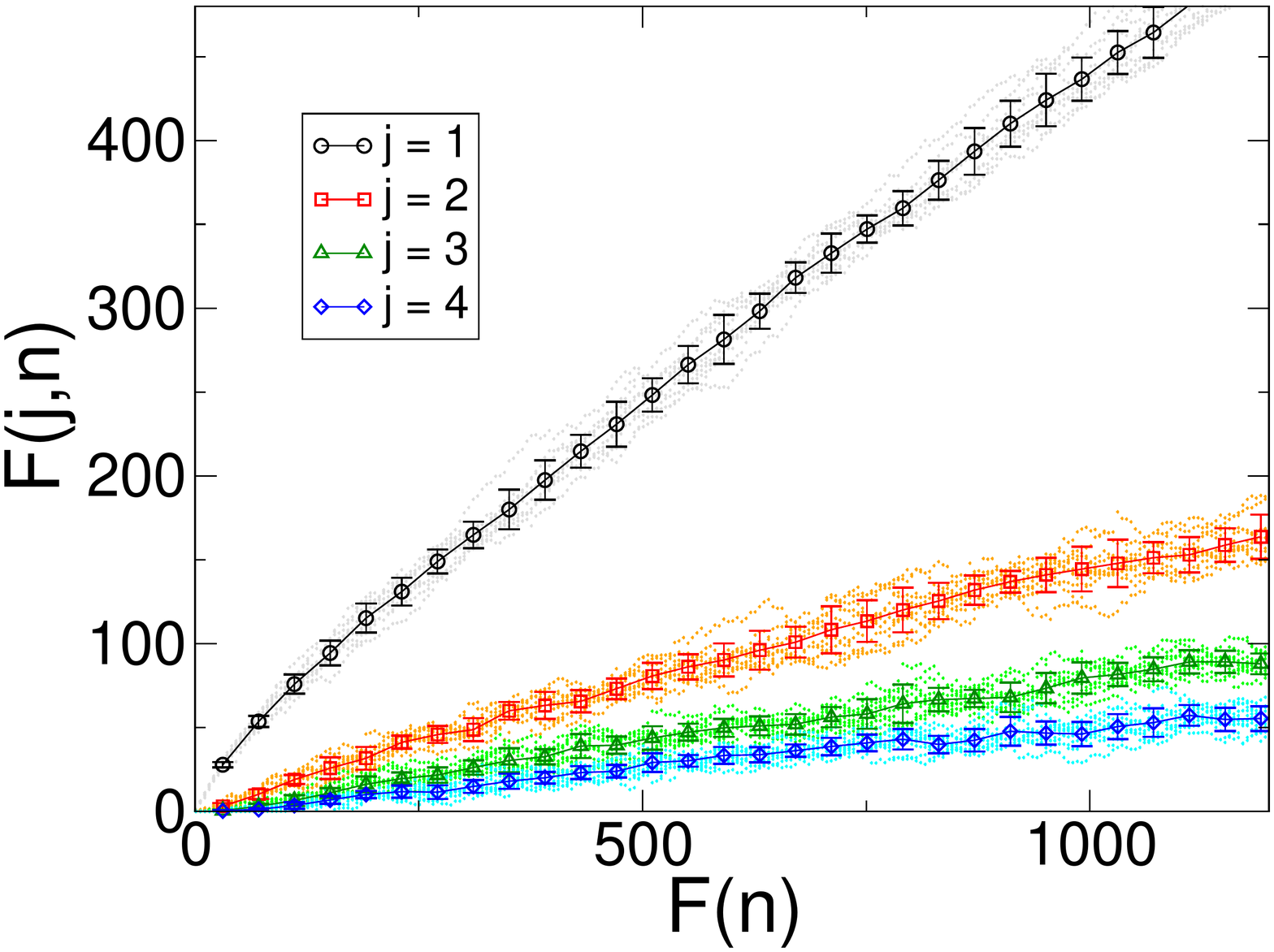}}
      \caption{(Color online) The classes with $j$ members form a finite
constant fraction of the total number of classes.  $F(j,n)$ (the number
of classes with $j$ members) is plotted versus $F(n)$ for different
values of $j$ in empirical data~\subref{sub:f1n_emp} and simulations of
the CRP ($\alpha = 0.32, \theta = 50$) without~\subref{sub:f1n_d0} and with
domain loss ($\delta = 0.2$) in model with uniform loss~\subref{sub:f1n_m1} and
weighted loss~\subref{sub:f1n_m2}. In the plots from simulations, the lines
in lighter colors represent 10 different realizations, while
the points with error bars are the mean value over them. The
figure shows that adding domain loss to the model has no
qualitative consequences in the behavior of the domain class
histograms.}
  \label{fig:F1n}
\end{figure}

It is possible to obtain the same results through a different route
compared to the above reasoning, by writing the hierarchy of
mean-field equations for $F(j,n)$, using a master equation-like
approach. Similarly as what happens for the Zero-Range
process~\cite{Evans05}, these equations contain source and sink terms
governing the population dynamics of classes.  Duplications create a
flux from classes with $j-1$ to classes with $j$ members, while only
$F(1,n)$ has a source term coming from the innovation move:

\begin{equation}
  \left \{  \begin{array}{ccc} 
      \partial_n F(n)=&\frac{ \a \:F(n) \:+ \theta}{n + \theta}\\
      \partial_n F(1,n) =&\frac{ \a \:F(n) \:+ \theta}{n + \theta} - (1-\a)
\frac{F(1,n)}{n+\theta}\\
      \partial_n F(2,n) =&(1-\a) \frac{F(1,n)}{n+\theta}- (2-\a)
\frac{F(2,n)}{n+\theta}\\
\cdots& \cdots
\end{array} \right . \ \ .
\end{equation}

We consider the limit of large $n$, and use the ansatz $F(j,n)= \chi_j
F(n)$.  This ansatz can be justified empirically and by simulations,
as shown in Fig.~\ref{fig:F1n}, which compares this feature in
empirical data and in simulations of different variants of the
model.

We have
\begin{equation}
 \left \{  \begin{array}{ccc} 
     \partial_n F(n)=&\frac{ \a \:F(n) \:}{n}\\
     \a \chi_1 =&\a-\: (1-\a)\chi_1 \\
     \a \chi_2 =& (1-\a) \chi_1- (2-\a) \chi_2\\
     \cdots& \\
     \a \chi_j =& (j-1-\a) \chi_{j-1}- (j-\a) \chi_j
\end{array} \right . \ \ ,
\end{equation}
giving
\begin{equation}
  \left \{  \begin{array}{ccc} 
      \chi_1 =&\a\\
      2\: \chi_2 =& (1-\a)\: \chi_1\\
      \cdots& \\
      j  \:\chi_j =& (j-1-\a)\: \chi_{j-1}
\end{array} \right . \ \ .
\end{equation}

The solution of these equations is
\begin{equation}
 \chi_j =\:\prod_{l=1}^{j-1} (l-\a) \frac{1}{\Gamma(j+1)}\: \a = 
\:\frac{\a}{\Gamma(1-\a)}\cdot (j-1)^{(1-\a)}\cdot
\frac{\Gamma(j-1)}{\Gamma(j+1)} \ \ ,
\end{equation}
which can be estimated as
\begin{equation}
 \chi_j = \a\:
\frac{1}{\Gamma(1-\a)}\:\left[\frac{1}{j}\right]^{1+\a} \ \ ,
\end{equation}
giving the result for the scaling of $F(j,n)$ and its prefactor.

\section{Direct simulation of the CRP and finite-size effects}

\begin{figure}[htbp]
   \centering
      \subfigure[]{%
        \label{sub:cum_hist}%
        \includegraphics[width=.45\textwidth]{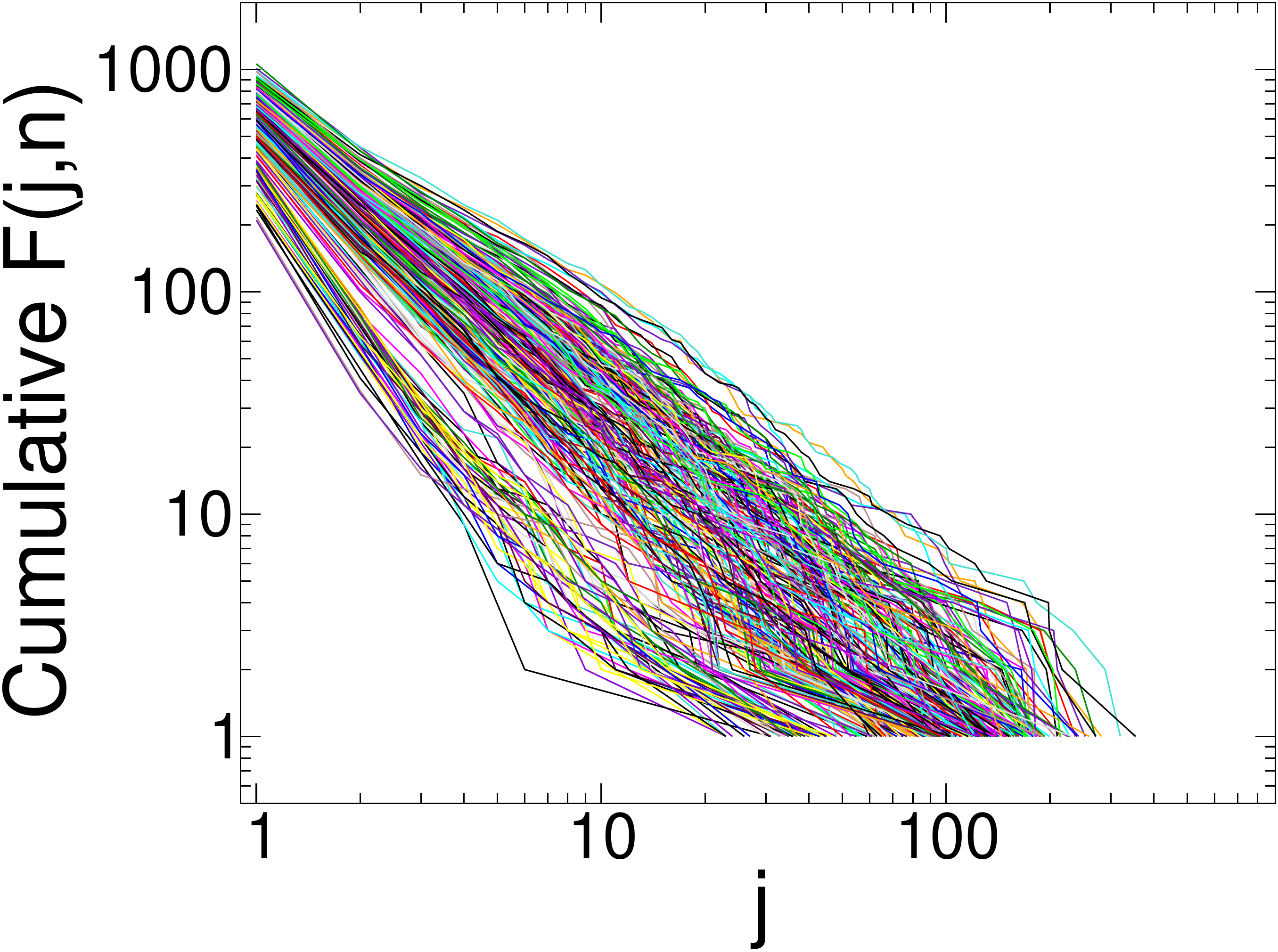}}%
\hspace{0.5cm}
      \subfigure[]{%
        \label{sub:alphafit}
        \includegraphics[width=.45\textwidth]{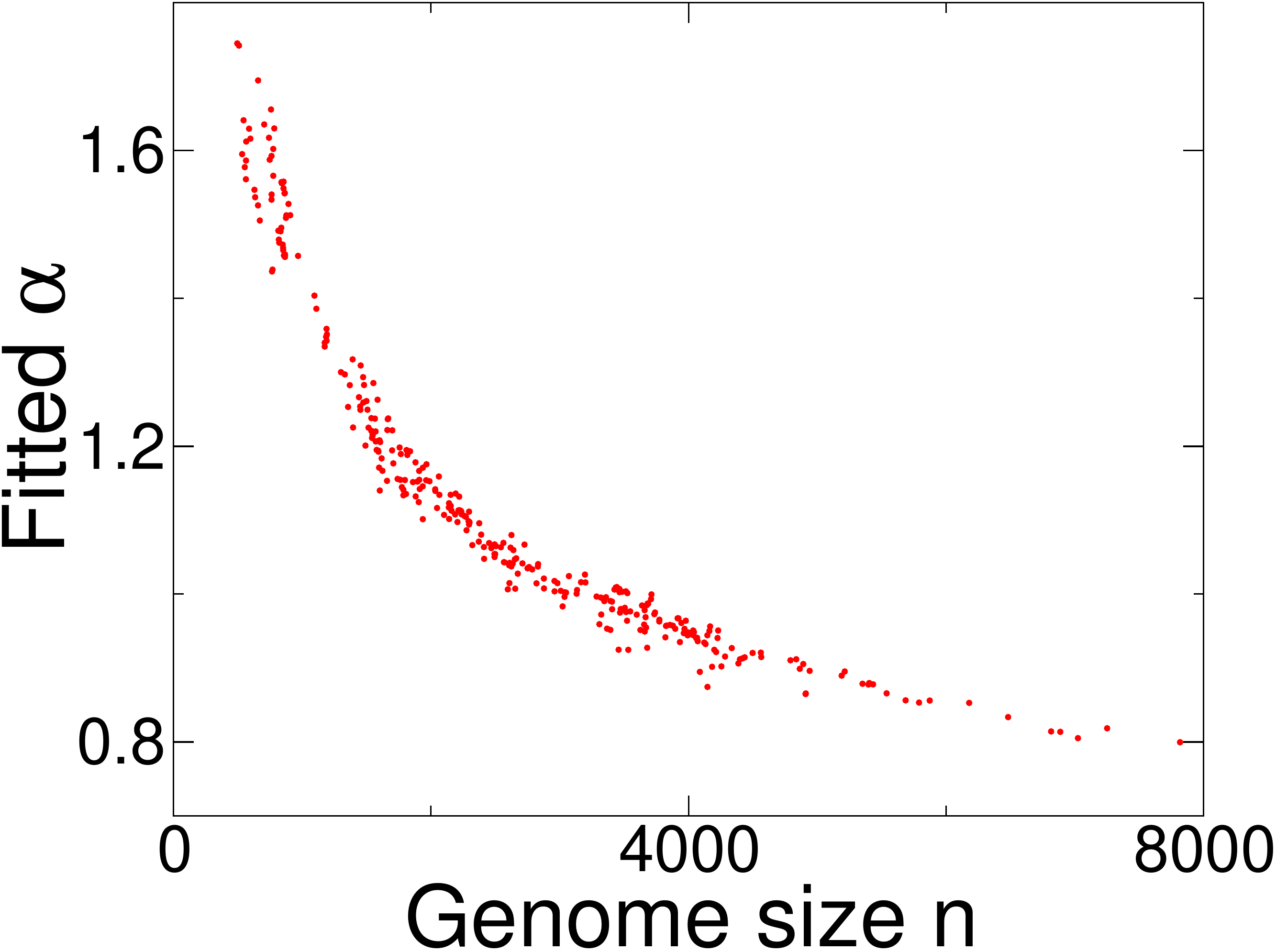}}%
      \caption{(Color online) \subref{sub:cum_hist} Cumulative distributions for
prokaryotes
        (data from the SUPERFAMILY database). Each is the cumulative
        histogram of domains in classes for a single genome.
        \subref{sub:alphafit} Fitted exponent of $F(j,n)$ from empirical data
        of all prokaryotes, plotted as a function of $n$.}
   \label{fig:alphafit}
\end{figure}

Going beyond scaling, the probability distributions generated by a CRP
contain large finite-size effects that are relevant for the
experimental genome sizes.  In this section we analyze the finite size
effect affecting the distribution over the domain classes, obtained
performing direct numerical simulations of different CRP realizations.
The simulations allow to measure $F(n)$, and $F(j,n)$ for finite
sizes, and in particular for values of $n$ that are comparable to
those of observed genomes shown in Fig.~\ref{fig:alphafit}.

\begin{figure}[htbp]
   \centering
   \subfigure[]{
        \label{sub:c1}%
        \includegraphics[width=.45\textwidth]{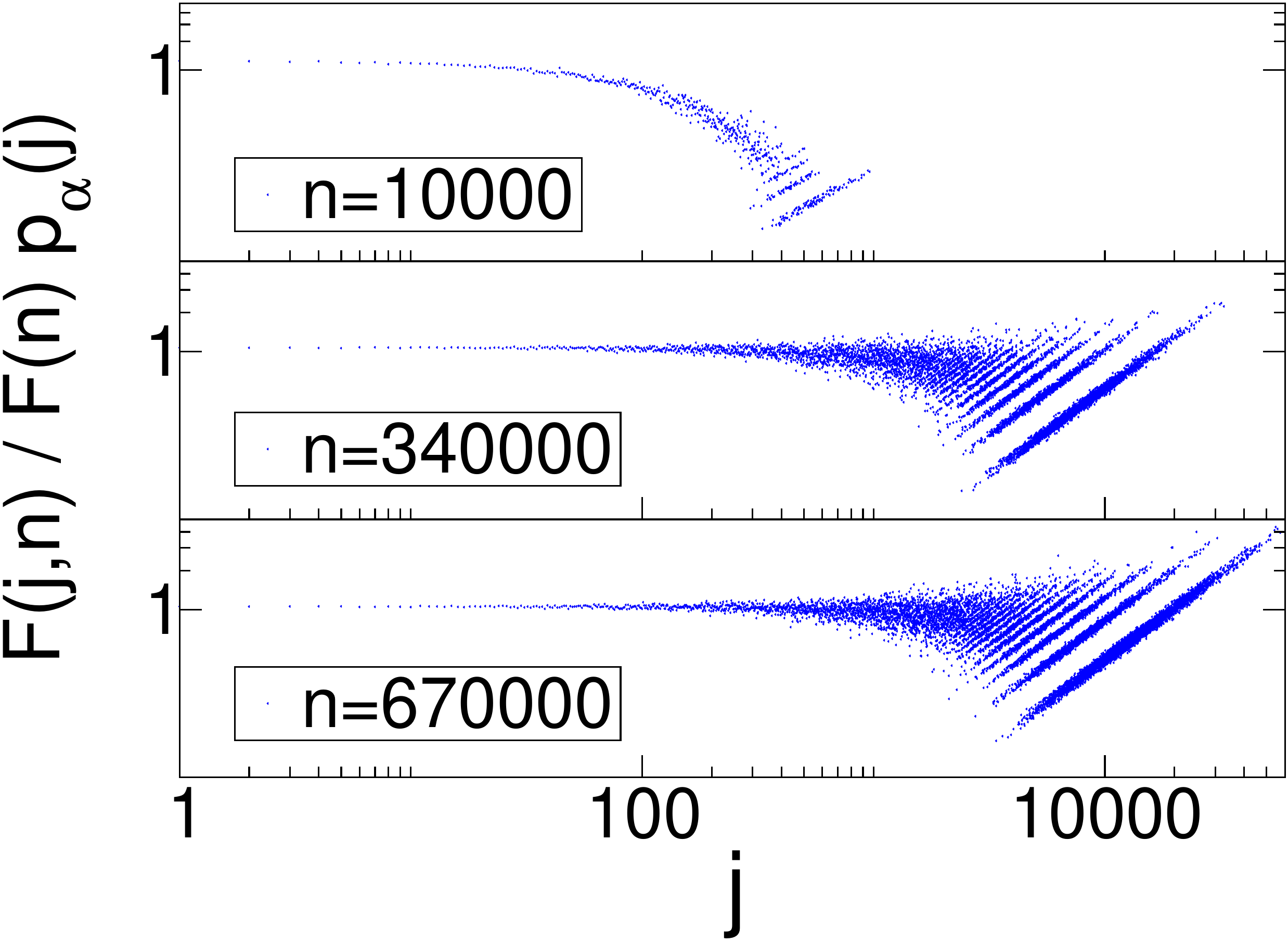}}
      \subfigure[]{%
        \label{sub:c2}%
        \includegraphics[width=.45\textwidth]{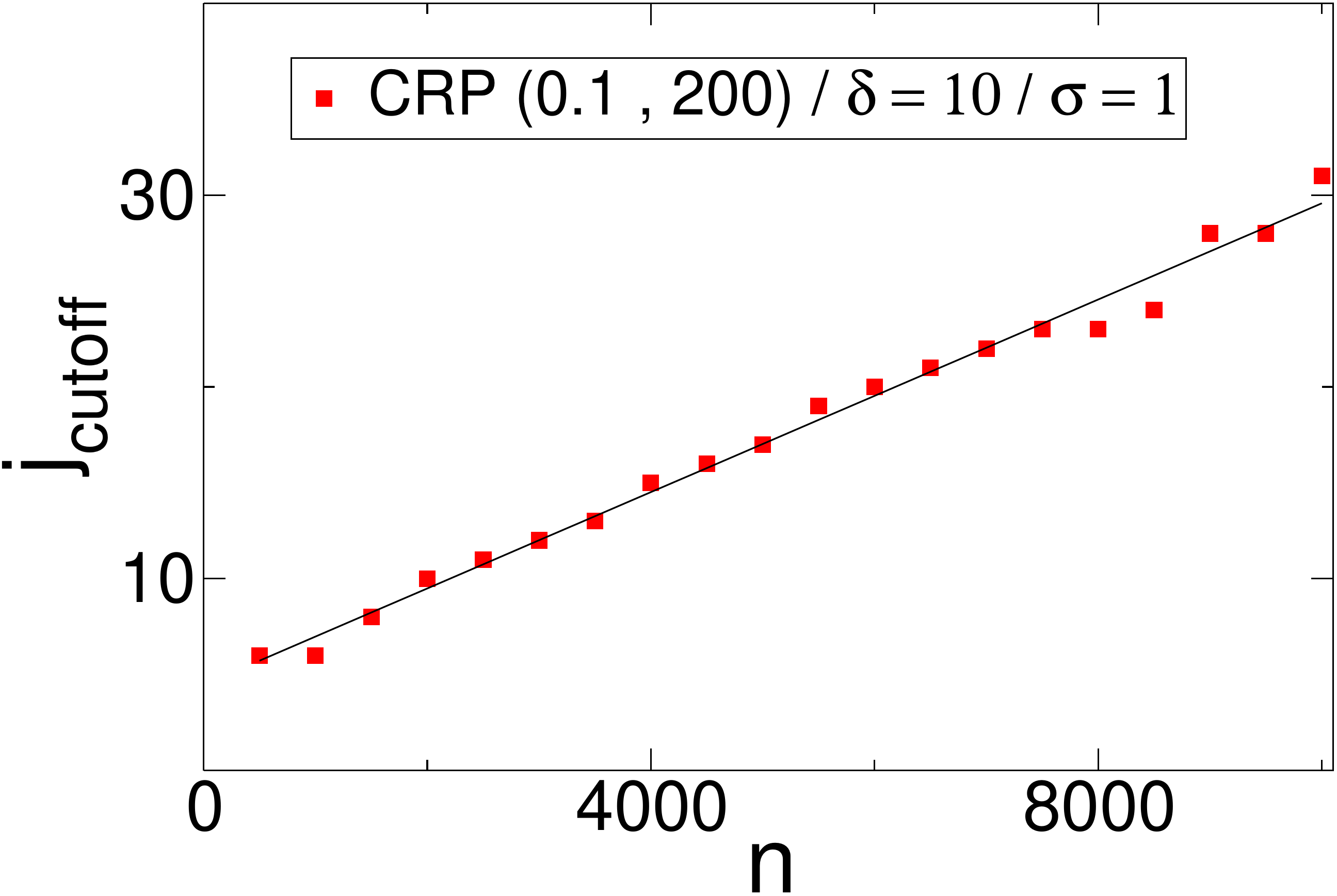}}%
      \subfigure[]{%
        \label{sub:cut_iter}
        \includegraphics[width=.45\textwidth]{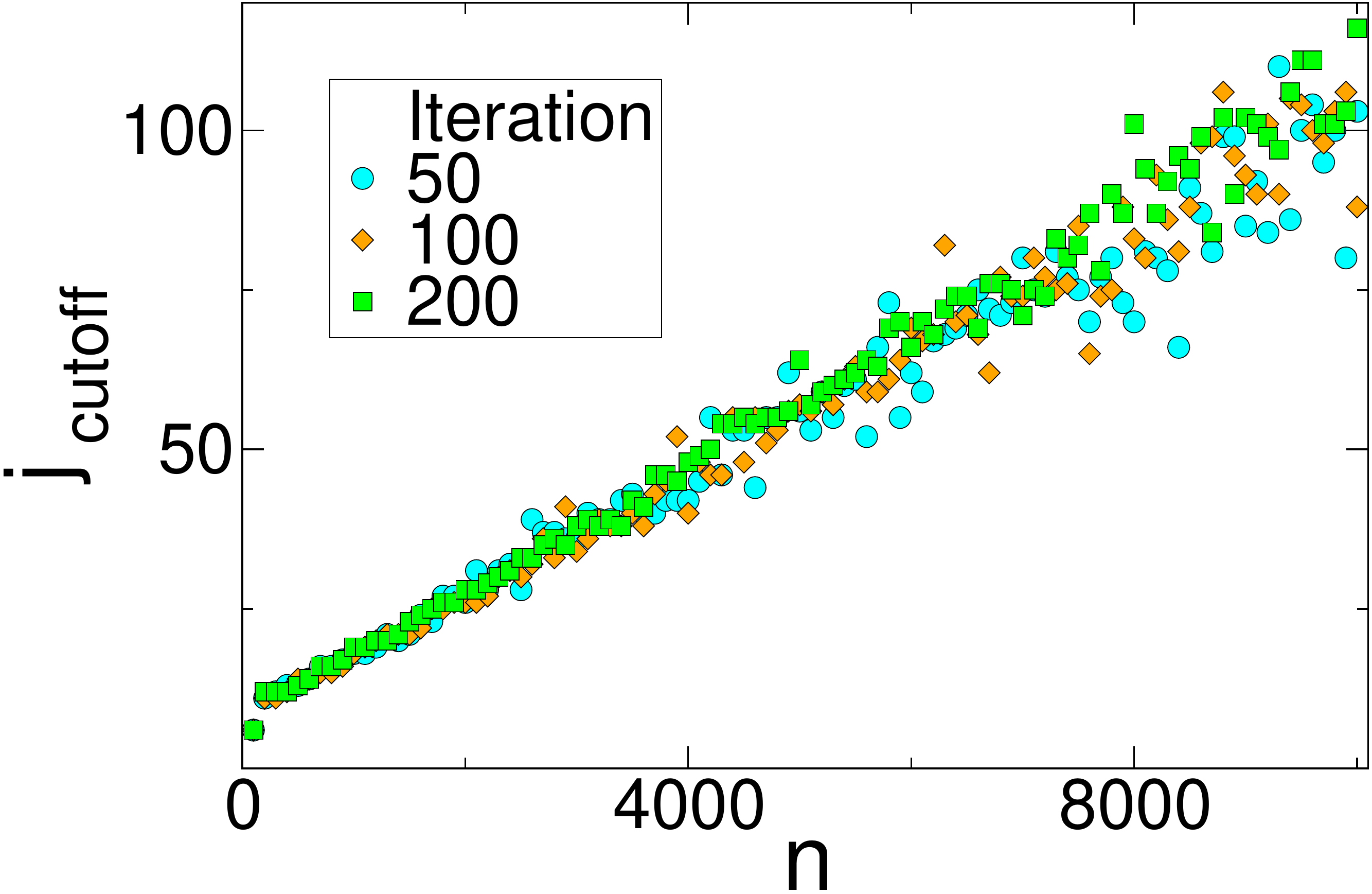}}%
      \caption{(Color online) \subref{sub:c1} Ratio
        of finite-size and asymptotic value for the distribution of
        domain classes population $F(j,n)$ taken from 500 realizations
        of a CRP with $\alpha=0.32$ and $\theta=50$. The three plots
        corresponds to different value of $n$. \subref{sub:c2} Linear
        $n$-dependency for the cutoff in a CRP. The parameters in the
        plot are $\alpha = 0.1$ and $\theta=200$. A similar trend can
        be observed for other values of the
        parameters. \subref{sub:cut_iter} Cutoff trends obtained from a CRP
        with $\alpha=0.44$ and $\theta=60$ and a different number of
        iterations. The procedure has parameters $K=10$ and
        $\delta=0.5$.}
   \label{fig:cutoff}
\end{figure}

The normalized distribution of the number of classes with $j$ domains
over a genome of length $n$ reaches the theoretical distribution
suggested by our model only in the asymptotic limit
\begin{equation}
  \frac{F(j,n)}{F(n)} \quad \stackrel{n \rightarrow \infty}{\longrightarrow}
\quad p_\alpha
  (j)=\frac{\alpha \Gamma(j-\alpha)}{\Gamma(1-\alpha)\Gamma(j+1)} \ .
\label{eq:an_as_crp}
\end{equation}
It is possible to obtain more information by studying the ratio of the
asymptotic distribution and the distribution obtained from the CRP
simulation as shown in Fig.~\ref{sub:c1}.  The plot shows that there is a
value $m^*(n)$, depending on the size $n$ of the genome, beyond
which the distribution is not anymore consistent with $p_\alpha(j)$
but shows exponential decay and large fluctuations.

In order to obtain a quantitative estimate of the deviation from
scaling generating the cutoff, we define an order parameter as
follows.  Using data obtained from the simulation as in Fig.~\ref{sub:c1},
we find the mean of the first 30 points of the plotted function and
then compute the standard deviation $\sigma$ by analyzing windows of
$K$ points together. The cutoff is defined when $\sigma > \delta$,
where $\delta$ is a parameter. The result of this procedure can be
seen in Fig.~\ref{sub:c2}.  The cutoff shows a linear dependence from the
genome length $n$. To make sure the procedure
does not depend too much on the number of iterations $N_I$ used to
obtain the mean value of $F(j,n)$, we performed it for different values of
$N_I$. As can be expected (Fig.\ref{sub:cut_iter}), more statistics is
needed for probing the cutoff trend in those regions where the
probability density function is very small.
Were it necessary to obtain from the distribution over domains $n$ an
estimate of parameters for the underlying CRP, one could decide to
consider only data with $j < j_{\text{cutoff}}$.
At the scales that are relevant for empirical data, finite-size
corrections are substantial.  Indeed, the asymptotic behavior is
typically reached for sizes of the order of $n \sim 10^6$, where the
predictions of the mean-field theory are confirmed.
Comparing the histogram of domain occurrence for the mean-field
solution of the model, simulations and data, it becomes evident that
the intrinsic cutoff set by $n$ causes the observed drift in
the fitted exponent of the empirical distribution visible in
Fig.~\ref{fig:alphafit}.  This means that the common behavior of the
slopes followed by the population of domain classes for genomes of
similar sizes can be ascribed to finite-size effects of a common
underlying stochastic process.

Beyond the linear cutoff, the behavior of the distribution becomes
realization-dependent due to the breaking of
self-average~\cite{condensation}. The relevant parameter to disentangle
the realization-dependence is $f$.  High-$f$ realizations have
different tails of the distribution from low-$f$ ones, giving rise to
the large fluctuations observed in Fig.~\ref{sub:c1}.
Thus, while the mean-field approach is successful in predicting the
asymptotic scaling of the distribution $F(j,n)$, it does not capture
the finite-size effects which can be observed in single realization of
the CRP process with finite $n$ and $f$.
Beyond mean-field is possible to obtain more information by
considering the sum of all CRP trajectories conditioned to reaching
configurations with given $n$ and $f$.

This enables a statistical-mechanical derivation of the normalized
distribution of the number of domains with $f$ classes over a genome
of length (number of domains) $n$. Since the focus here is on the mean-field
approach, the calculation is described in a parallel work~\cite{condensation}.

\section{Occurrence of domain classes and CRP realizations}

The above model does not describe evolutionary time in
generations. Conversely, it reproduces random ensembles of different
genomes generated one from the other with the basic moves of
duplication, innovation (and loss, see below). It considers only
events that are observed at a given $n$, independently on when or why
they happened in physical or biological time.
Genomes from the same realization can be thought of as a trivial tree
of life, where each value of $n$ gives a new specie. In the case
including domain deletions, more genomes of the same history can have
the same size. In contrast, independent realizations are completely
unrelated.

The scaling laws in $F(n)$ and $F(j,n)$ hold for the typical
realization, indicating that the scaling laws originate
from the basic evolutionary moves and not from the fact that the
species stem from a common tree with intertwined paths due to common
evolutionary history. For example, two completely unrelated
realizations will reach similar values of $F$ at the same value of
$n$.

The data confirm this fact: phylogenetically distant bacteria with
similar sizes have very similar number and population distribution of
domain classes (see Fig.~\ref{fig:crpVSpwl}).

While the scaling laws are found independently on the realization of
the Chinese restaurant model, the uneven occurrence of domain classes
can be seen as strongly dependent on common evolutionary history.
Averaging over independent realizations, the prediction of the CRP is
that the frequency of occurrence of any domain class would be equal,
as no class is assigned a specific label. In the Chinese restaurant
metaphor, the customers only choose the tables on the basis of their
population, and all the tables are equal for any other feature.

\begin{figure}[htb]
  \centering
  \includegraphics[width=0.8\textwidth]{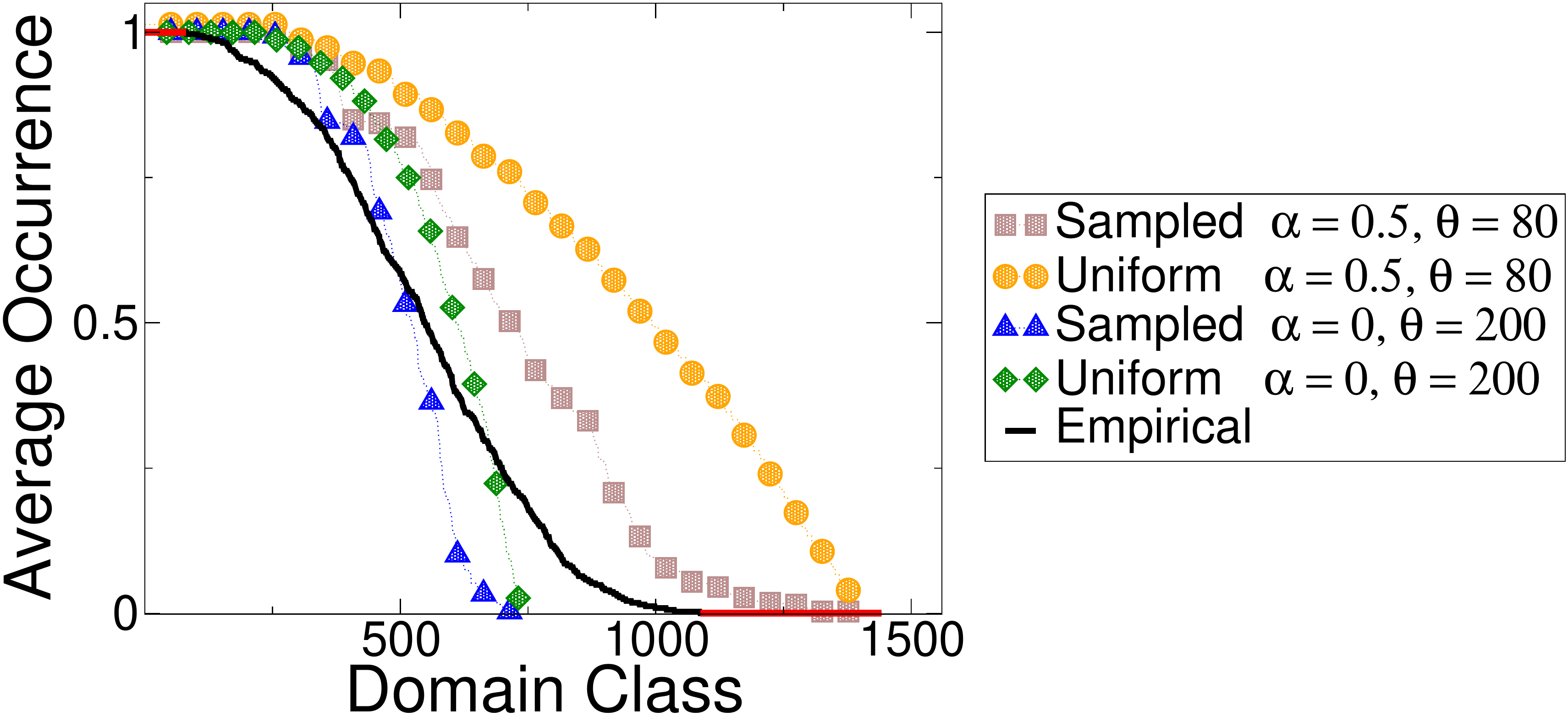}%
  \caption{(Color online) Single realization averages reproduce qualitatively
    empirical topology occurrence (the fraction of genomes where the given
domain
    topology is found, normalized by the total number of genomes). Domain
topology ranked occurrence
    in typical realizations of the CRP with different parameters
    (symbols), compared to the empirical data for bacteria (continuous
    line). The
    simulated data were obtained considering the occurrence curve for
    500 realizations of the process for all sizes from $n=300$ to
    $n=8000$ (``uniform''), i.e.\ the range of sizes observed in the
    data, or directly for the set of empirical sizes of the genomes
    (``sampled''). }
\label{fig:occurrence}
\end{figure}

In order to capture this behavior with the model, one can consider
the statistics of domain topology occurrence of a single realization,
which is an extremely crude, but comparatively more realistic
description of common ancestry. In other words, in this case, the
classes that appear first are obviously more common among the genomes,
and the qualitative phenomenology is restored, without the need of any
adjustment in the model definition (Fig.~\ref{fig:occurrence}).

\section{Domain Loss}

Loss of genes, and thus of domains, is reported to occur frequently in
genomes.  We will discuss now variants of the model considering the
introduction of a domain deletion, or loss rate.  The question we ask
is whether the introduction of domain loss, which we consider mainly
as a perturbation, affects the qualitative behavior of the model, for
example by generating different scaling behavior or phase transitions.
We will see that the answer to all these questions is mostly negative even
for non-infinitesimal perturbations, provided the loss rate is
constant and does not scale itself with $F$ and $n$. The main
exception to this behavior is found when the loss probability of a domain
depends on its own class size.

We introduce domain loss through a new parameter $\delta$, which
defines a loss probability $p_{L}$ in two \emph{a priori}
different ways. (1) We can distribute this probability equally among
domains, so that the per-class loss probability is $p_L^i =
\delta \frac{k_i}{n}$. Consequently, the duplication and innovation
probabilities $p_O$ and $p_N$ are rescaled by a factor $(1-\delta)$.
(2) We can also weigh the loss probability of a domain on its own
class size, in the same way as domains are duplicated in the standard
CRP, so we obtain a per-class loss move with probability $p_L^i
= \delta \frac{k_i - \alpha}{n + \theta}$, giving a total $p_L =
\delta \frac{n - \alpha F}{n + \theta} $ and the rescaling of $p_O$
and $p_N$.
We will see that model (1) and (2) are not equivalent.

On technical grounds, the introduction of domain loss makes the
stochastic process entirely different: $n$ is now a random variable,
and all the observables that depend on it (e.g.\ $F(n)$) are stochastic
functions of this variable. Another parameter, $t$, describes the
iterations of the model.
Operatively, we tackle the two models with the usual mean-field
approach, writing equations for $n$ and $F$ of the kind $\partial_t n
= \textsc{q}(n,F)$, $\partial_t F = \textsc{r}(n,F)$, and hence obtain
the behavior as a function of $n$ by considering $\partial_n F =\frac{
  \textsc{r}(n,F)}{\textsc{q}(n,F)}$.
The exact meaning of these equations is not straightforward.  For
example $F(n)$ should represent the average on all histories passing
by $n$, but the differential equation strictly describes only the
dependence of the observable from the actual value of the random
variable $n$. Nevertheless, the predictions of this mean-field
approach agree well with the results of simulations, indicating that
these complications typically do not affect the behavior of the means.
We will consider situations where, on average, genomes are not
shrinking.

Considering model (1), we can write the mean-field equations as
\begin{equation}
  \partial_t F(t) = (1 - \delta) \frac{\alpha F(t) + \theta}{n + \theta} -
\delta \frac{F(1,n)}{n}
\end{equation}
\begin{equation}
 \partial_t K_{i}(t) = (1 - \delta) \frac{K_{i}(t) - \alpha}{n + \theta} -
 \delta \frac{K_{i}(t)}{n} \ \ ,
\end{equation}
where the sink term for $F$ derives from domain loss in classes with a
single element, quantified by $F(1,n)$. Since time does not count
genome size, one has to consider the evolution of $n$ with time $t$,
given in this case by $\partial_t n = 1 - 2 \delta$.
In order to solve these expressions, we use the ansatz $F(1,n) =
\chi_1 F(n)$, and considering the limit in large $n$.  The ansatz is
verified by simulations and holds also for empirical data, as
previously shown (Fig.~\ref{fig:F1n}). The first equation reads:

\begin{equation}
  \frac{\partial_{n}{F(n)}}{F(n)} = \frac{1}{n} \left[ \frac{(1-\delta)\alpha
   -\delta\chi_1}{1-2\delta} \right] \ \ .
\end{equation}

\begin{figure}[htbp]
   \subfigure[]{
        \label{sub:c1a}
        \includegraphics[width=.45\textwidth]{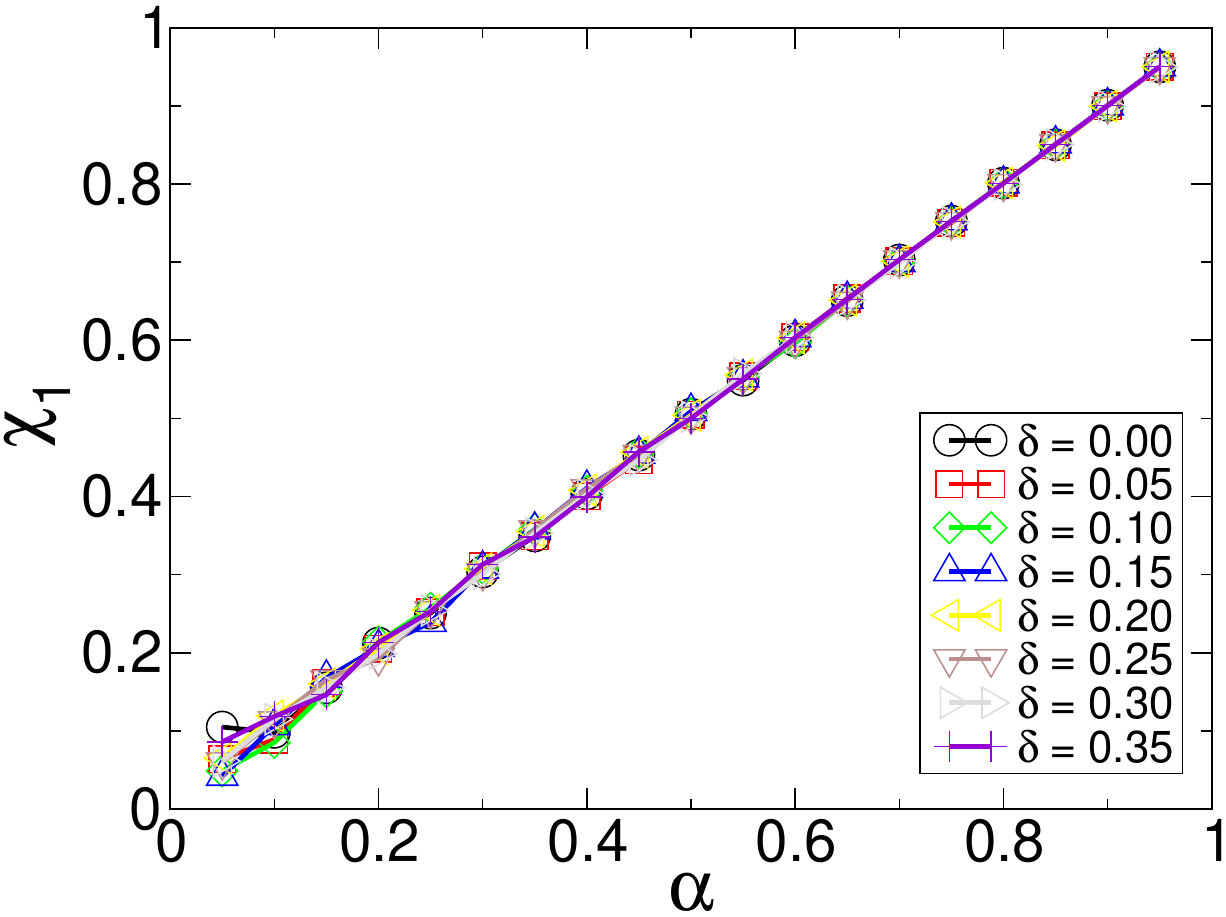}}
      \subfigure[]{
        \label{sub:ara}
        \includegraphics[width=.45\textwidth]{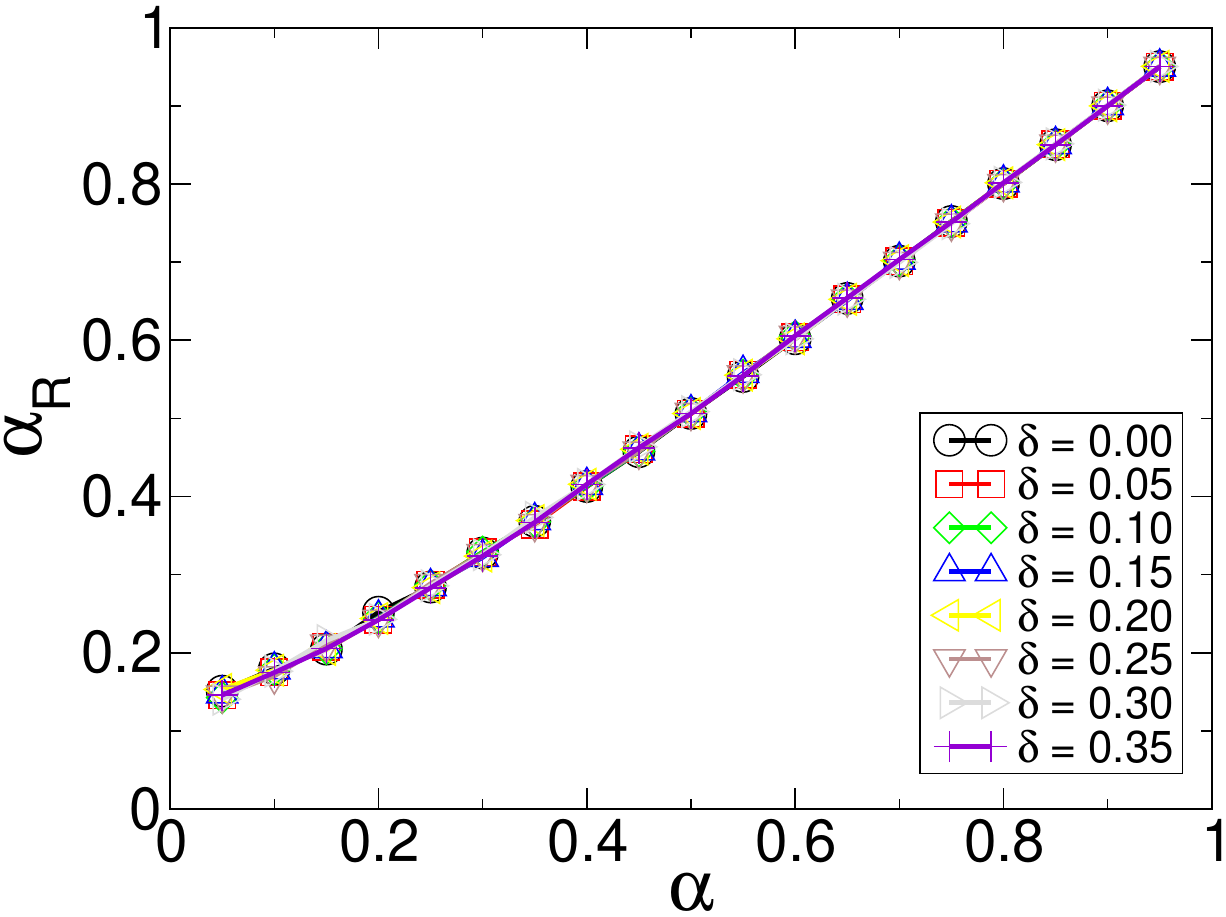}}
      \subfigure[]{
        \label{sub:an}
        \includegraphics[width=.45\textwidth]{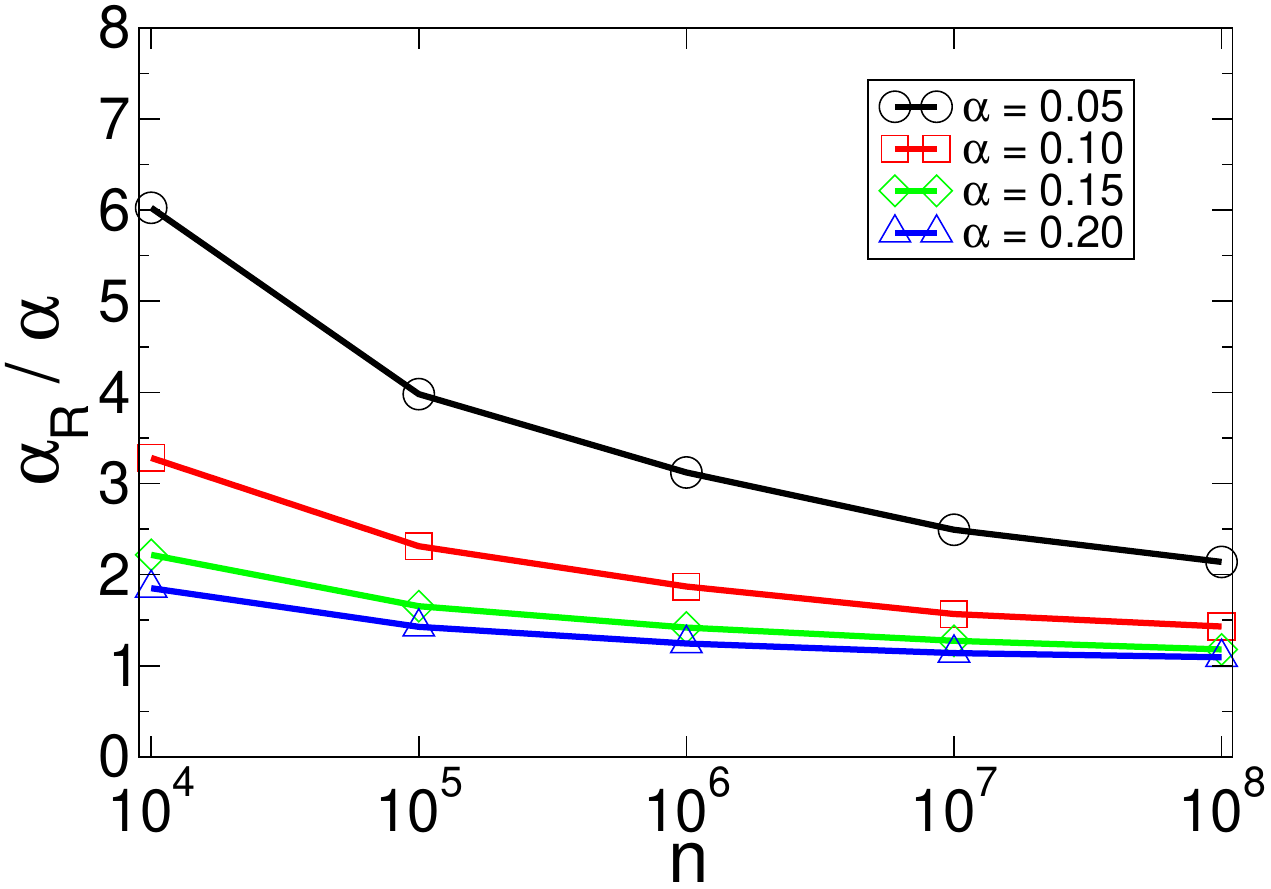}}
      \subfigure[]{
        \label{sub:qrd}
        \includegraphics[width=.45\textwidth]{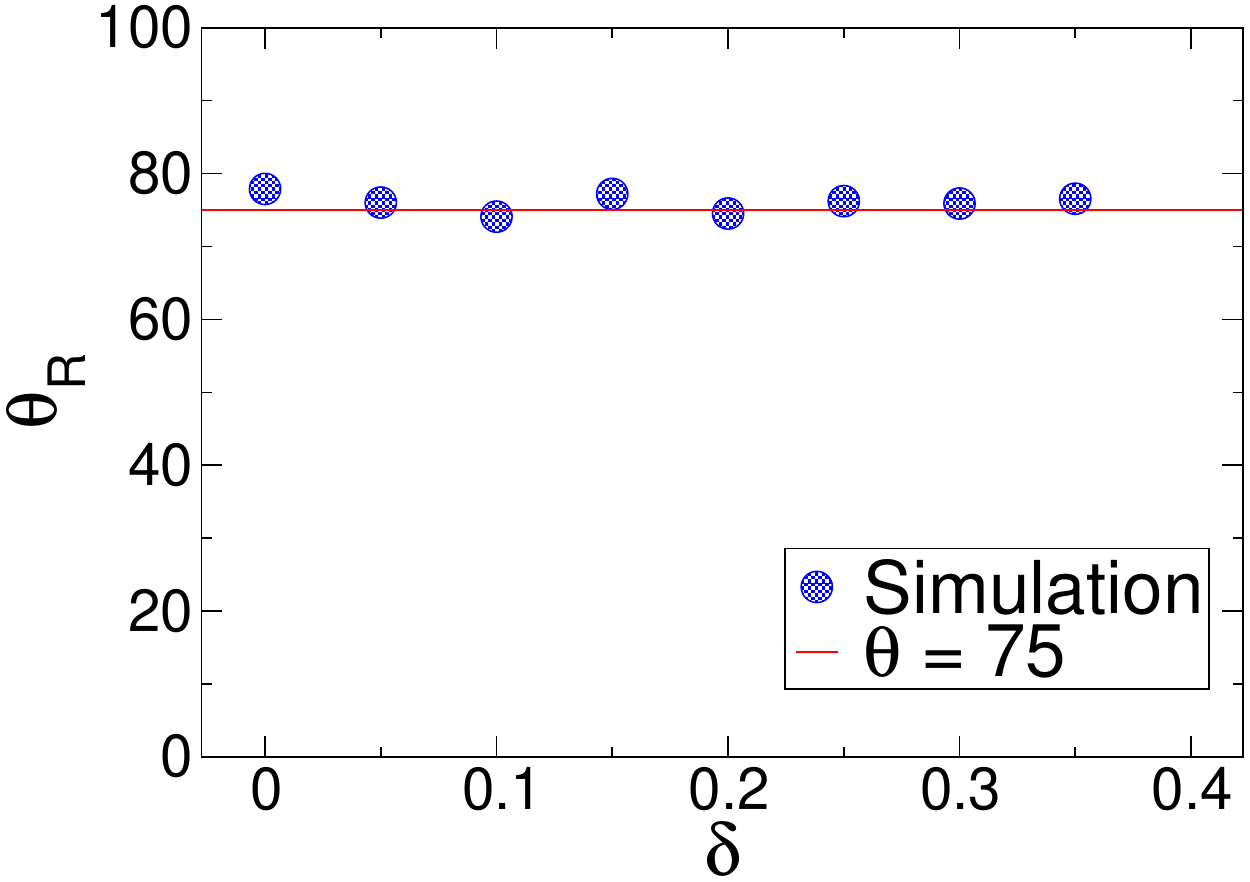}}
      \caption{(Color online) A model with uniform domain loss (model (1)) does
not
        lead to a change of qualitative behavior with respect to
        absence of domain loss. We estimated $\chi_1$ from simulations
        with a linear fit of the measured (linear) $F(1,n)$.  The
        measured coefficient $\chi_1$ \subref{sub:c1a} approaches with the
        parameter $\alpha$, hence the observed scaling exponent
        $\alpha_R$, measured from the simulated $F(n)$ agrees very
        well with the imposed $\alpha$. \subref{sub:ara} This trend is weaker
for
        smaller $\alpha$ but can be regarded as a finite size effect,
        as the agreement improves \subref{sub:an} for increasing $n$. The plot
in
        (c) refers to $\theta = 75, \d=0.35$. Finally \subref{sub:qrd}, the same
        phenomenology holds for $\alpha = 0$, where the observed
        parameter $\theta_R$ extracted from the behavior of $F(n)$ in
        simulations agrees well with $\theta$.}
  \label{fig:loss_unif}
\end{figure}

The above equation gives the conventional scaling for $F(n)$ and
$K(n)$ with $\alpha$ replaced by $\alpha_R = \frac{(1-\delta)\alpha
   -\delta\chi_1}{1-2\delta}$, the correction resulting from the measured value
of
$\chi_1$.
By the use of computer simulations we notice that the
$\chi_1$ coefficient tends to $\alpha$ for infinite-size genomes
(Fig.~\ref{sub:c1a}, \ref{sub:ara}, and \ref{sub:an}), so
that the asymptotic trend of the equally distributed domain loss is
identical to that of the standard CRP. This
behavior is independent from the chosen $\delta < 1/2$, as the
asymptotic regime depends only on the growth of $F(n)$, governed by
$\alpha$.

When we analyze the domain distribution of finite-size genomes, we
obtain the conventional results: the power-law depends on the genome
size, but not on the value of $\delta$ (Fig.~\ref{sub:ara} and \ref{sub:qrd}).
This is explained considering the fact we are comparing runs with
fixed genome size, thus with different number of moves, and we do not
consider genomes that lose all their own domains. In fact biologically
one cannot trace the number of moves needed to reach a specific
genome, but essentially we can  observe only genomes in their actual
state.

More precise results can be obtained by the use of the mean field
``master equation'' approach sketched above. Using the same ansatz  
$F(j,n) = \chi_j F(n)$, we obtain the following hierarchy of equations
for $\chi_j$
\begin{equation}
  [B + \d j + (1-\d)(j-\a)] \chi_j=(1-\d)(j-1-\a)\chi_{j-1}+\d
  (j+1)\chi_{j+1} \ \ ,
\end{equation}
with $B = (1-\d)\a-\d \chi_1$ . It is
possible to estimate the solution of this system by taking a continuum
limit as
\begin{equation}
B\chi(x) +(1-\d)\partial_x [(x-a)\chi(x) ]- \d\partial_x \chi(x)=0 \ \ ,
\end{equation}
which can be solved giving
\begin{equation}
  \chi(x) \sim \left[ \frac{1}{x} \right] ^{1+ Z} \ \ ,
\end{equation}
with $Z = \frac{B}{1-2\d}$. We also find $F(n) = n^Z$, which is
consistent with the constraint $\sum_j\: j \chi_j= \frac{n}{F(n)}$
since $\sum_j\: j \chi_j \sim n^{1-Z}$.
It is then clear that the introduction of domain loss is equivalent to
a rescaling of the parameter $\a$ to $ \a_R = Z(\a,\d)$, but in our
case, $Z(\a,\d) = \a$ asymptotically. 

The case $\a = 0$ has to be treated separately, but
the behavior is similar (Fig.~\ref{sub:qrd}).

\begin{figure}[htbp]
   \centering
   \subfigure[]{
        \label{sub:aramfa2}
        \includegraphics[width=.4\textwidth]{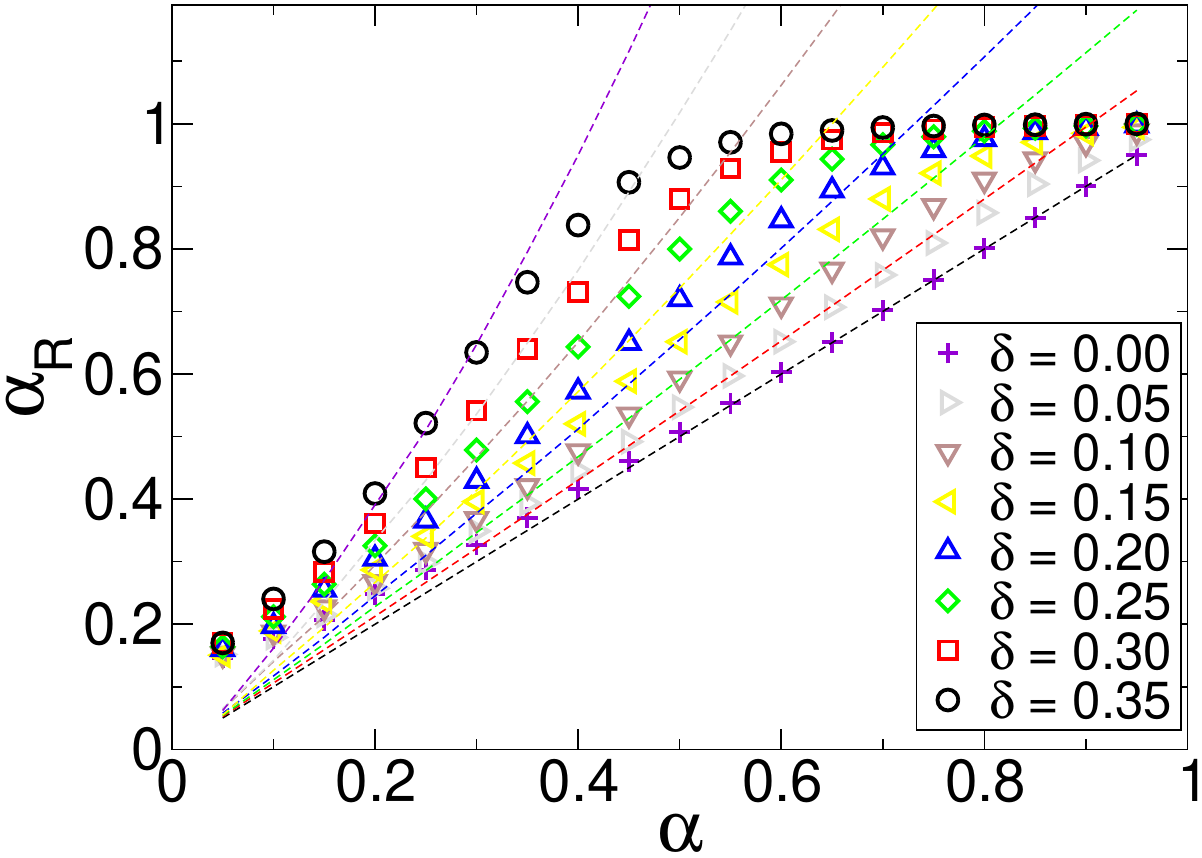}}
      \subfigure[]{%
        \label{sub:fase}
        \includegraphics[width=.45\textwidth]{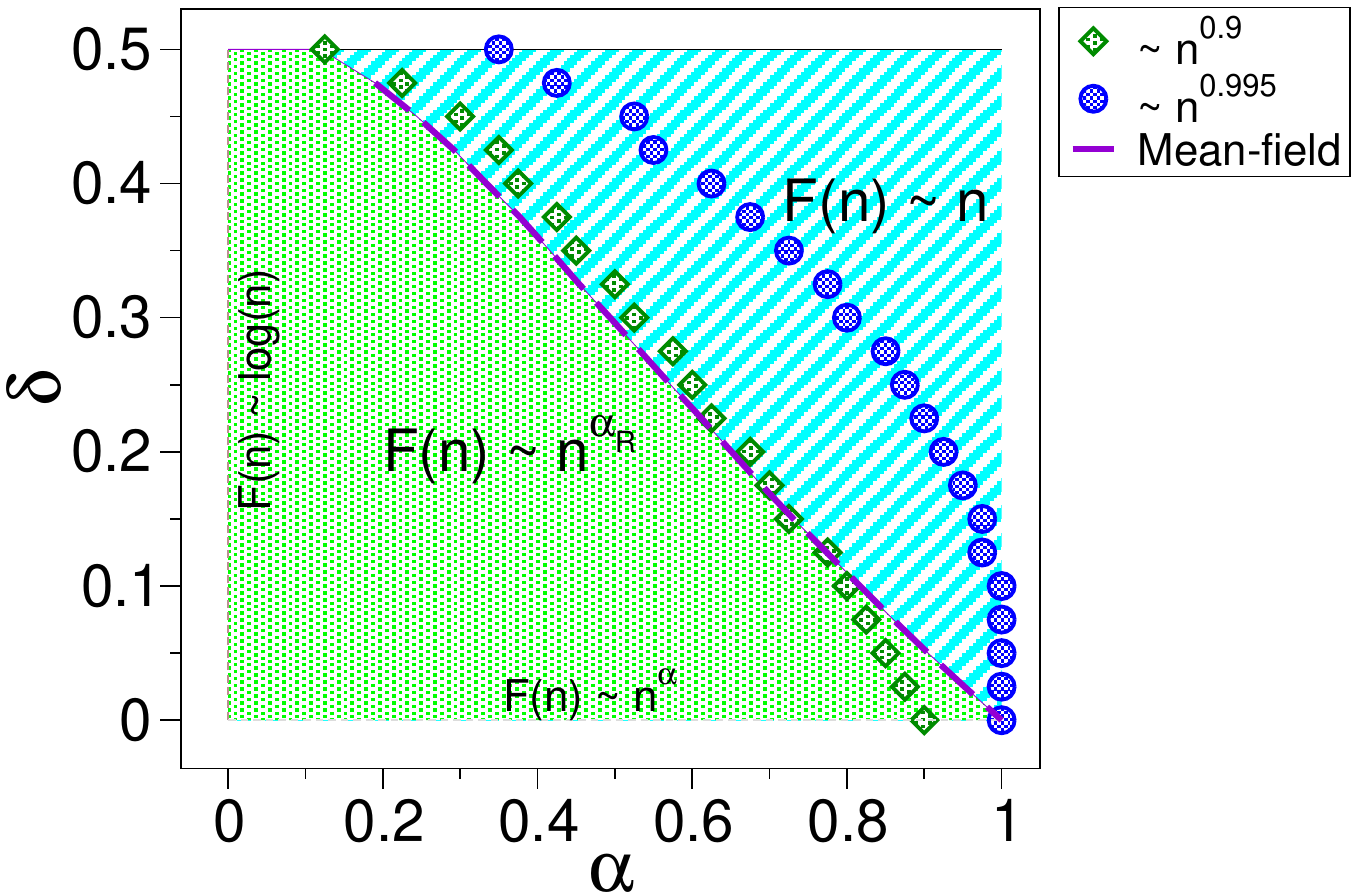}}%
      \caption{(Color online) In a model with weighted domain loss (model (2)),
the
        loss probability can trigger a transition between two
        different scaling behaviors for $F(n)$.  \subref{sub:aramfa2} behavior
of
        observed exponent $\alpha_R$ versus $\a$ in simulations. The
        exponent saturates to unity at a critical value that is lower
        than one. The dashed lines are mean-field predictions in the
        ansatz of $F(n)$ sublinear in $n$. \subref{sub:fase} ``Phase-diagram''
for
        the behavior of $\a_R$ for different values of $\a$ and
        $\d$. The dashed line is a mean-field estimate of the
        transition point to $\alpha_R = 1$, while circles and diamonds
        are the $\a_R=.9$ and $\a_R=.995$ lines from simulations at
        $n=10^6$.    }
  \label{fig:loss_weigh}
\end{figure}

A similar procedure is applicable to model (2). In this case, however,
the dependence of the effective death rate from $n$ can bring to an
interesting change in the phenomenology, where $\d$ can select the
observed exponent $\alpha_R$ and also determine a regime of linear
growth for $F(n)$.

To understand this point we can consider the mean-field evolution of
$F(n)$.  This is determined asymptotically by the balance of a growth
term $p_N \sim \a F/n$ and a loss term $p_L \sim \d F(1,n) (1-\a)/n
$. With the usual ansatz this gives an asymptotic evolution equation
of the kind
\begin{equation}
  \partial_n F(n)= F(n) \frac{\a - \d(1-\a)\chi_1} {(1- 2\d)n  + 2 \d
    \a F(n)} \ ,
  \label{eq:mortepes}
\end{equation}
with the usual definitions.  Simulations confirm the ansatz $F(1,n) =
\chi_1 F(n)$ for this second model,
which we will use in the mean-field reasoning.  If $F(n)$ is intensive,
i.e.\ asymptotically of order inferior to $n$, the term $2\d\a F$ can
be neglected and this equation gives the scaling law $F \sim
n^{\alpha_R}$, with $\alpha_R = \frac {\a - \chi_1 \d (1-\a)}{1-2\d}$.
However, this equation has solution also if $F$ is order $n$,
i.e.\ $\a_R=1$, and the term $2 \d \a F$ in Eq.~(\ref{eq:mortepes})
cannot be neglected. In this case $\a$ and $\d$ determine the
prefactor of the scaling law $F \sim n$.

Thus, there are two self-consistent mean-field asymptotic solutions,
and we expect a transition between the two distinct behaviors. The
existence of this transition is confirmed by simulations
(Fig.~\ref{sub:aramfa2}): at fixed $\d$, $\a_R$ saturates to $\a_R \simeq 1$ for
larger values of this
parameter. The transition point can be understood in mean-field as the
intersection of the two solutions $\a_R = 1$ and $\a_R = \frac {\a -
  \chi_1 \d (1-\a)}{1-2\d}$ at varying $\d$, and gives rise to a
two-parameter ``phase-diagram'' separating the linear from the
sublinear scaling of $F(n)$ with $n$, as shown in
Fig.~\ref{sub:fase}.

In conclusion, the mean-field approach is effective in exploring the
effects of domain loss, which, under some general hypotheses, does not
disrupt the basic phenomenology of the duplication-innovation model.
Specifically, there appear to be no qualitative changes introduced by
a finite uniform loss rate, as long as this rate is constant with
$n$. A loss rate that is weighted as the innovation rate, instead, can
induce an interesting transition from sublinear to linear scaling.

Thinking about the empirical system, no direct
quantitative estimates are currently available regarding the domain
loss rate as a function of genome size or number of classes. For this
reason, it currently appears difficult to make a definite choice for
this ingredient in the model.

\section{Models with domain class specificity}
\label{sec:CRPfit}

In the previous sections we analyzed models that make no distinction
between domain topologies, but the latter are selected for duplication
moves only on the basis of their population.

It is then clear that they can reproduce the observed qualitative trends
for the domain classes and their distributions with one common set of
parameters for all genomes.  One further question is to estimate the
quantitative values of these parameters for the data.  While the
empirical slope of $F(n)$ could be seen as more compatible with a
model having $\alpha=0$, as its slope decays faster than a power law
for large values of $n$, the slopes of the power-law distribution of
domain classes $P(j,n)$ and their cutoff as a function of $n$ is in
closer agreement to a CRP with $\alpha$ between 0.5 and
0.7~\cite{Lagomarsino2009}.

We will now discuss a CRP variant that is able to distinguish domain classes
based on \emph{a priori} information, thus breaking the symmetry of the model
by exchange of tables.

This is equivalent to the
introduction of differently colored ``tablecloths'' labeling the
tables, that are determinant in the choice for one table or the
other.

Those table colors can be set by any observable of interest. In our
analysis, we considered the empirical occurrence of a domain topology
as a label. Indeed, the occurrence of a given domain class is determined
by its biological function. For example, as expected, all the ``core''
biological
functions such as translation of proteins and DNA replication are
performed by highly occurring domain topologies, since this machinery
must be present in each genome. Accordingly, these universal classes
performing core functions have to appear preferentially earlier on in
a model realization.
This variant of the model is important for producing informed null
models for the analysis of the empirical data, and, as we will see,
shows the best agreement with respect to the scaling laws.

\begin{figure}[htbp]
	\centering
	\includegraphics[width=.85\textwidth]{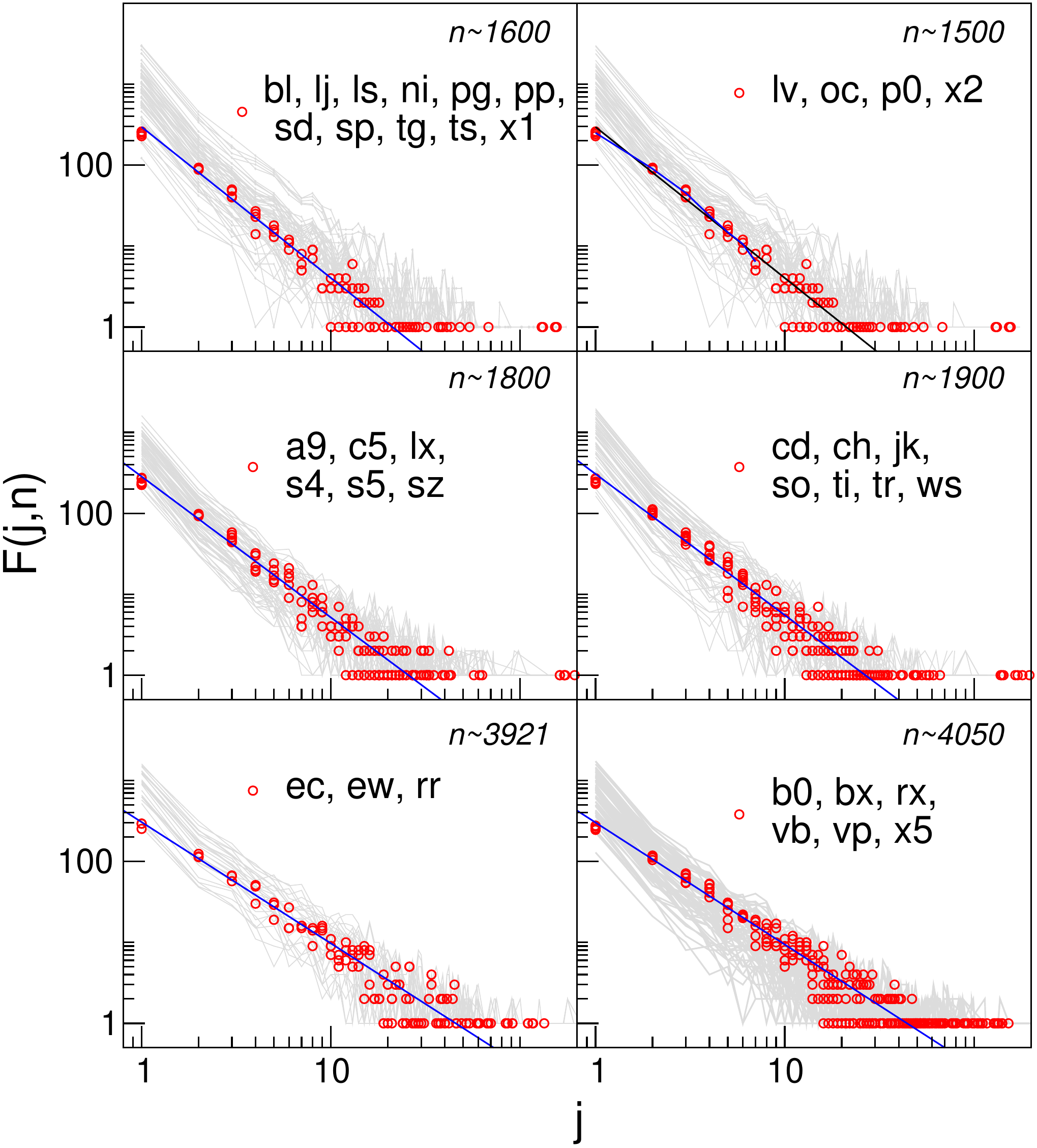}
        \caption{(Color online) Comparison of $F(j,n)$ from the
          SUPERFAMILY Database with mean-field predictions and
          simulations of the model with class specificity.  Comparison
          between $n-$dependence of powerlaw fit (lines in blue) and
          universality of the two parameters from our model. Two
          letters identify each genome, whose full name can be found
          in Appendix (Table~\ref{GenName}).  Simulations (lines in
          gray) from our model use the same values $\alpha=0.75$ and
          $\theta=32$.  }
	\label{fig:crpVSpwl}
\end{figure}

We will introduce the variant with class specificity by coupling the
CRP model to a simple genetic algorithm able to select between
innovation moves that choose different classes.  Let us first
introduce some notation to parametrize domain class occurrence. We
define the matrices $\sigma_i^k$ where:
\begin{equation}
 \sigma_i^g := \left\{ \begin{array}{ll}
 	1 & \textrm{domain $i$ found in genome $g$}\\
 	-1 & \textrm{domain $i$ not found in genome $g$}
 \end{array} \right.
\end{equation} 
It is possible to consider the mean taken along the matrix columns,
\begin{equation}
  \left\langle \sigma_i^{\textrm{emp}} \right\rangle =\frac{1}{N^g} \sum_g
\sigma_i^{g,\textrm{emp}} \ \ ,
\end{equation}
where the label ``emp'' means that this value is obtained from
empirical data (Fig.~\ref{fig:occurrence}).

Generically, a genetic algorithm requires a representation of the
space of solution $\Omega$ and a function $f(\omega)$ that tests the
quality of the solution computed. In our case, the former is simply
the genome obtained from the a CRP step, parametrized by
$\sigma_i^k$. The latter is defined as
\begin{equation}
 \mathcal {F}(g)=\prod_i \exp\left( \, \sigma_i \left\langle
\sigma_i^{\textrm{emp}}\right\rangle\right)  = \exp\left( \, \sum_i \sigma_i
<\sigma_i^{\textrm{emp}}>\right) \ \ .
\end{equation} 
The value of the above scoring function taken over simulated genomes measures
how much the set of domain classes they possess agrees with
experimental data (in this case on occurrence) and enables to compare
different ``virtual'' CRP moves.

We consider the case of two virtual genomes $g^{'}$ and $g^{''}$
generated through standard CRP steps, for simplicity without domain
loss, from an initial size $n$ and genome $g(n)$.

Note that in this variant, since the empirical domain topologies are a
finite set, domain classes are also finite. As a consequence, tables
with a given tablecloth are extracted without replacement, affecting
the pool of available colors. As we will see, this is an important
requirement to obtain agreement with the data, as it determines the
saturation of the function $F(n)$ also for large values of $\a$.  We
will discuss the role of an infinite pool in the following subsection.

Also, as anticipated, domain classes have different ``color'', or in
mathematical terms the exchangeability of the process is lost. Classes
are drawn from the set of the residual ones with uniform probability. Genomes
$g^{'}$ and $g^{''}$ are compared through the function $\mathcal F$
and the highest score one will be the genome $g(n+1)$ so that $g(n+1) =
\mathrm{argmax}(\mathcal{F}(g^{'}),\mathcal{F}(g^{''}))$.

In these conditions, the rigorous results present in the literature for
the CRP cease to be valid.
It is still possible, however, to analyze the behavior of this
variant by the mean-field approach adopted here, and to compare with
simulations.
Since the selection rule chooses strictly the maximum, it is
essentially able to distinguish the sign of $\langle
\sigma^{\textrm{emp}}_i \rangle$ only. For this reason, it is
sufficient to account for the positivity (which we label by ``+'') and
negativity (``-'') of this function for a given domain index $i$.
This means that, with the simplification of two virtual moves only,
the model introduces only one extra effective parameter, i.e.\ the
ratio of the ``universal'' (positive $\mathcal{F}$) to the
``contextual'' (negative $\mathcal{F}$) domain classes.

In order to write the mean-field equations for this model variant, we
first have to classify all the possible outcomes of the virtual CRP
moves.  The genomes $g^{'}$ and $g^{''}$ proposed by the CRP
proliferation step can have the same (labeled by ``$1$''), higher
(``$1_+$'') or lower (``$1_-$'') score than their parent, depending
on $p_O$, $p_N$ and by the probabilities to draw a universal or
contextual domain family, $p_+$ and $p_-$ respectively.  Using these
labels, the scheme of the possible states and their outcome in the
selection step is given in Table~\ref{tab:moves}.

\begin{table}[htbp]
\begin{center}
\begin{tabular}{|c|c|c|}
    \hline
    proliferation $(g^{'},g^{''})$ & probability & selection \\
    \hline
    $(1,1)$& $p_O^2$ & old \\
    $(1,1_{-})$ & $2\: p_O\:p_N\:p_-$ & old \\
    \hline
    $(1,1_{+})$ & $2\: p_O\:p_N\:p_+$ & new+ \\
    $(1_{+},1_{+})$ & $p_N^2\:p_+^2$ & new+ \\
    $(1_{+},1_{-})$ & $2\:p_N^2\:p_-\:p_+$ & new+ \\
    \hline
    $(1_{-},1_{-})$ & $p_N^2\:p_-^2$ & new- \\
    \hline
\end{tabular}
\end{center}
\caption{Compound probabilities of picking up a new domain or an old domain with
positive or negative cost function.}
\label{tab:moves}
\end{table}

From the Table~\ref{tab:moves}, it is straightforward to derive the modified
duplication and innovation probabilities $\hat p_O$ and $\hat p_N$ of
the complete algorithmic iteration:
\begin{equation}
  \hat p_O= p_O \:(p_O+ 2\:p_N p_-)
\end{equation}
\begin{equation}
  \hat p_N= p_N \:(p_N+ 2\:p_O p_+) = p_{N+} + p_{N-} \ \ ,
\end{equation}
where $ p_{N+} = p_N p_+ (2 - p_N p_+)$ and $ p_{N-} = p_N^2 (1 - p_+)^2 $ 
are the probabilities that the new domain is drawn from the universal
or contextual families respectively.

We now write the macroscopic evolution equation for the number of
domain families by the usual procedure. Calling
$K^+(n)$ and $K^-(n)$ the number of domain classes that have positive
or negative $ \langle \sigma^{\textrm{emp}}_i \rangle$ and are
not represented in $g(n)$,

\begin{equation}
 \left \{  \begin{array}{cc} 
 \partial_n F(n)=& \hat p_N\\
\partial_n K^{+}(n)=& - \hat p_{N+}\\
\partial_n K^{-}(n)=& - \hat p_{N-}\\
\end{array} \right . \ \ .
\end{equation}

Now, $p_+ = K^+ / (K^- + K^+) = K^+ / (D - F(n))$, so that we can
rewrite 

\begin{equation}
 \left \{  \begin{array}{cc}
 \partial_n F(n)=& (\frac{\alpha F(n)+\theta}{n+\theta})\:\Big[
\frac{\alpha F(n)+\theta}{n+\theta}\:+ \: \frac{2
  K^{+}(n)}{D-F(n)}(\frac{n-\alpha F(n)}{n+\theta})\Big]\\ 
\partial_n K^{+}(n)=&-(\frac{\alpha F(n)+\theta}{n+\theta})\:  \frac{
  K^{+}(t)}{D-F(n)}\:\Big[2-\: (\frac{\alpha F(n)+\theta}{n+\theta}) 
 \frac{K^{+}(n)}{D-F(n)}\Big]\\
\partial_n K^{-}(n)=& - (\frac{\alpha F(n)+\theta}{n+\theta})^2\:
(\frac{ K^{+}(n)}{D-F(n)})^2 
\end{array} \right . \ \ .
\label{eq:din}
\end{equation}

The above equations have the following consistency properties
\begin{itemize}
\item $\partial_n \Big(K^{+}+ K^{-}+F\Big)=0$, hence $K^{+}+ K^{-}+F =
  D\quad \forall n$. 
\item $\partial_n F \leq 1$, hence $ F(n) \leq n$. 
\item $\partial_n F \geq 0$, $\partial_n K^+ \geq 0$ and $\partial_n
  (F + K^+)  \geq 0$ so that $F$ grows faster than $K^+$ decreases.  
\end{itemize}
Choosing the initial conditions from empirical data $n_0, F(n_0)$, size
and number of domain classes of the smallest genome, we have, since
$F(n_0) < n_0$ and $\alpha \le 1$, 
\begin{equation}
  \frac{ \alpha F(n_0) + \theta } { n_0 + \theta } < 1 \ \ .
\end{equation}
It is simple to verify that under this condition the system always has
solutions that relax to a finite value $F_{\infty} < D $. Indeed, after the
time $n^*$ where $K^+(n*) = 0$, the equations reduce to $\partial_n
K^+ = 0$, $K^- = D- F$ and 
\begin{equation}
  \partial_n F(n)=\left(\frac{\alpha F(n)+\theta}{n+\theta}\right)^2 \ \ ,
\end{equation}
immediately giving our result. 
It is important to notice that this result depends on the fact the
empirical reservoir of domains is finite (and thus on the biological
hypothesis that the full pool of available domain topologies is).
Indeed, in presence of an infinite reservoir of domain classes, $F(n)$
does not relax to a finite value. In this case, similarly as in the
case of non uniform loss, there is a transition from linear to
sublinear scaling piloted by the parameters $p_+$ and $\a$. If
$2p_+\a<1$, $F\sim n^{2p_+\a}$, while if $2p_+\a>1$, $F\sim n$.

\begin{figure}[htbp]
  \centering \includegraphics[width=0.7\textwidth]{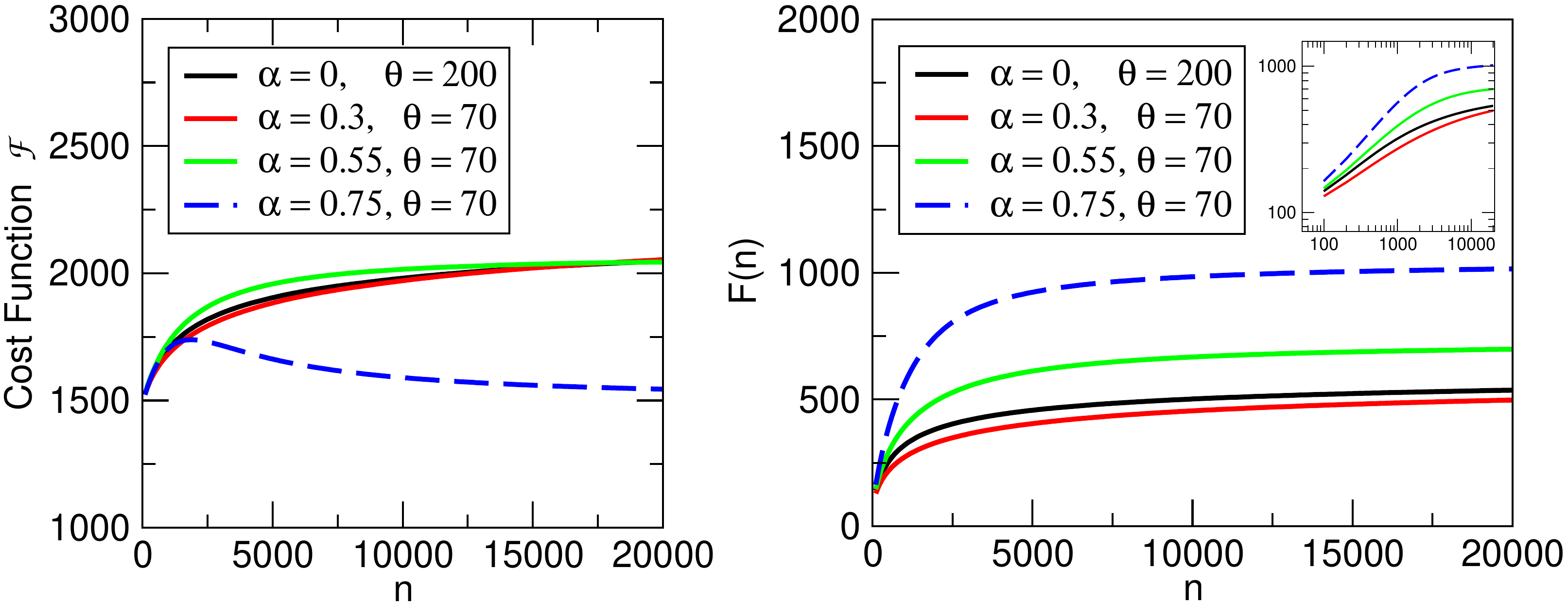} 
  \caption{(Color online) Numerical solutions of the mean-field equations of the
CRP
    model with selection of specific domain classes. Left panel: score
    function $\mathcal{F}(n)$ for different values of $\alpha$. Right
    panel: $F(n)$ plotted in linear and logarithmic (inset) scales.}
  \label{fig:num_fitness}
\end{figure}

\begin{figure}[htbp]
   \centering
        \includegraphics[width=0.4\textwidth]{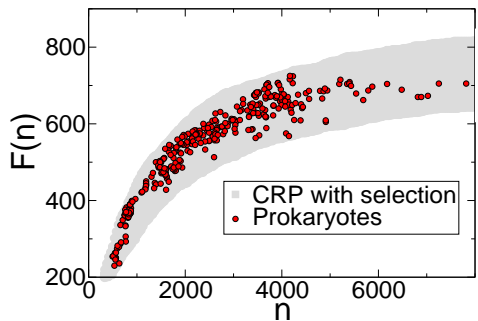}
	\caption{(Color online) Comparison of the CRP with class specificity
with
          experimental data. CRP parameters are $\alpha=0.75$ e
          $\theta = 0.32$. Number of domain classes versus the length
          $n$ of genome.  Simulations were performed for $n < 10000$
          taking data at each interval of $1000$ units in size, and
          considering 5 realizations at each step.}
   \label{fig:agreement}
\end{figure}

Numerical solutions of Eq.~(\ref{eq:din}) give the same behavior for
$F(n)$ as the direct simulations (Fig.~\ref{fig:num_fitness}). In
particular, while this function grows as a power law for small genome
sizes, it saturates at the relevant scale, giving good agreement with
the data (Fig.~\ref{fig:agreement}). The internal laws of domain usage of this
model were obtained from direct simulations only, and give a good
quantitative agreement with the data (Fig.~\ref{fig:agreement}).
Finally, Fig.~\ref{fig:num_fitness} also shows that, for large values of
$\alpha$ (above $0.7$) this function reaches a maximum at sizes
between 2000 and 4000. This is also where most of the empirical
genomes are found, indicating that this range of genome sizes may
allow the optimal usage of universal and contextual domain families.

\section{Conclusions}


We presented from a statistical physics viewpoint a class of
duplication-innovation-loss stochastic processes able to describe the
probability distributions and scaling laws observed in genomic data
sets for protein structural domains with few effective parameters.
These models are different declinations of the basic paradigm set by
the Chinese restaurant process which, though much explored in the
statistical literature, remains relatively unexplored by the physics
community, despite of its rich and peculiar phenomenology, which could
make it useful in multiple applications.

Our focus has been to present the basic phenomenology of different
variants of the model connected to possibly relevant aspects of the
evolutionary dynamics of protein domains, prominently domain
duplication, innovation, loss, and the specificity of domain classes.
In doing this, we have shown how a mean-field approach, despite of its
simplicity, can be extremely powerful in the analysis of the
qualitative behavior of this class of models.
More subtle aspects of these models might be approached by simulations
and refined statistical mechanics methods, directly accessing the sum
on all different paths.

For the standard CRP, we have shown how the scaling laws in the
number and in the slope of the observed power-law distributions of
domain classes are qualitatively reproduced, by the
typical or the average realization.
The salient ingredient for this feature is that this model can include the
correct relative scaling of innovation to duplication moves, which can
be generated by statistical dependence of innovation, or by the fact
that the process of domain birth does not give rise to identically
distributed random variables at different sizes.

The mean-field results can be obtained in two independent ways: with
dynamic equations for the population and number of domain classes, or,
similarly as for the Zero-range process~\cite{Evans05}, by considering
the evolution of the number of classes with a given population. The
latter ``master equation'' approach gives access to the full histogram
of domain class population.
The scaling of the slope of the domain class distributions can be
understood as a finite-size effect on these histograms.
We have also shown how the uneven occurrence of domain classes can be
explained by the CRP provided one considers the statistics of
occurrence in a single realization. This can be interpreted as the
fact that genomes are related by common evolutionary paths.

This phenomenology is extremely robust to the introduction of finite
domain loss probabilities that do not scale with $n$ or $F$. For
uniform domain loss, the full mean-field equations are still
accessible analytically, and point to the effect of domain loss as a
simple effective rescaling of the model parameters. The presence of a
loss rate that scales with size, instead, can trigger a transition
from sublinear to linear scaling in the number of distinct classes.

Finally, we have discussed the case of models that introduce the
specificity of domain classes, and thus explicitly breaks the symmetry
between them, as expected for the biological case. Such variants can
be very useful in more detailed statistical investigations (e.g.\ as a
null model) and to define inference problems on the available
bioinformatic data.  

For example, a preferential duplication of conserved proteins has been
reported in eukaryotes~\cite{DP04}, and could be
tested against a duplication-innovation-loss model where a
``conservation index'' characterizes specificity.

Here, we have shown how a CRP variant with
specificity can be formulated as a genetic algorithm where different
CRP virtual moves are selected on the basis of an informed scoring
function, defined on the basis of further empirical observables
related to domain classes (such as function, correlated occurrence,
etc.)
In our case, we have shown how such a variant, weighting the virtual
moves according to the observed occurrence of the finite pool of
domain classes has very good quantitative agreement with the available
data.

The future perspective on this systems and models are abundant, both
for the application of the models to the genomics of protein domains,
where the most promising ways seems to be the use in specific inference
problems on species with known evolutionary histories, and in other
problems of statistical physics and complex systems, for example to
develop new growth models for complex networks where the nodes can evolve with
similar moves.
One prominent example is the case of protein-protein interaction
networks, graphs where the nodes follow a dynamics that is coupled to
those of protein domains, while the edges can be inherited by
duplication, lost, or rewired by mutation and natural selection.

\begin{acknowledgments}
  GB acknowledges financial support from the project IST STREP
  GENNETEC contract No. 034952.
\end{acknowledgments}

\newpage

\appendix

\setcounter{table}{0} 
\renewcommand{\thetable}{\thesection.\arabic{table}}

\section{Data and methods}

All the data considered here were extracted from the SUPERFAMILY
database version 1.69 and 1.73. We considered domains at the
superfamily level, although similar results hold for folds and
families. A genome ``size'' $n$ was defined as the number of domain
hits for each genome. The main data structures considered were the
number of domain classes and the histograms of domain class
population. We also considered the ranked occurrence of a given domain
class across genomes.

The different variants of the duplication-innovation-loss model were
simulated directly using a C++ code and compared with the mean-field
results and the empirical data. The variant with class specificity
couples the CRP model to a simple genetic algorithm selecting
innovation moves that choose distinct classes, on the basis of their
empirical occurrence. For all the variants, we derived and solved the
mean-field equations for the evolution of the main observables, and
described alternative approaches for their solution.

The finite-size behavior of a CRP was studied comparing the
finite-size histograms obtained from direct simulations with the
asymptotic limit of the exact analytical result (formula
\ref{eq:an_as_crp})~\cite{Pit02}, and defining a cutoff as the class-size where
the
deviation was larger than an arbitrary threshold. The cutoff was
studied as a function of $n$. For the CRP variants with specificity
and domain loss only the mean-field solution given here is available
for the same analysis. The exponent of the empirical domain class
distributions for genomes with different size was estimated from a
power-law fit of the cumulative histograms, keeping into account the
finite-size cutoff.

\clearpage

\section{List of Bacteria Used in the Analysis of Fig.~\ref{fig:crpVSpwl}}

\setstretch{0.8}

\begin{table}[htbp]
{\footnotesize
\begin{tabular}{c|c}
\textbf{Abbreviation }& \textbf{Full Name}\\ \hline \hline
a9	&	Streptococcus agalactiae A9	\\
b0	&	Bacillus anthracis Sterne	\\
b1	&	Baumannia cicadellinicola Hc	\\
bl	&	Bifidobacterium longum NCC2705	\\
bx	&	Bacillus cereus ATCC 10987	\\
c5	&	Chlorobium chlorochromatii CaD	\\
cd	&	Corynebacterium diphtheriae NCTC 13129	\\
ch	&	Chlorobium tepidum TLS	\\
ec	&	Escherichia coli K12	\\
ew	&	Escherichia coli W3110	\\
jk	&	Corynebacterium jeikeium K411	\\
lj	&	Lactobacillus johnsonii NCC 533	\\
ls	&	Lactobacillus sakei ssp. sakei 23K	\\
lv	&	Lactobacillus salivarius ssp. salivarius UCC118	\\
lx	&	Leifsonia xyli ssp. xyli CTCB07	\\
ni	&	Neisseria gonorrhoeae FA 1090	\\
oc	&	Prochlorococcus marinus NATL2A	\\
p0	&	Candidatus Protochlamydia amoebophila UWE25	\\
pg	&	Porphyromonas gingivalis W83	\\
pp	&	Streptococcus pyogenes MGAS2096	\\
rr	&	Rhodopseudomonas palustris BisB5	\\
rx	&	Rhodoferax ferrireducens T118	\\
s4	&	Streptococcus agalactiae 2603V/R	\\
s5	&	Streptococcus agalactiae NEM316	\\
sd	&	Streptococcus thermophilus LMG 18311	\\
so	&	Synechococcus sp. CC9605	\\
sp	&	Streptococcus pyogenes MGAS10394	\\
sz	&	Synechococcus sp. CC9902	\\
tg	&	Streptococcus pyogenes MGAS10750	\\
ti	&	Thiomicrospira denitrificans ATCC 33889	\\
tr	&	Thermus thermophilus HB8	\\
ts	&	Streptococcus pyogenes MGAS10270	\\
vb	&	Vibrio vulnificus YJ016	\\
vp	&	Vibrio parahaemolyticus RIMD 2210633	\\
ws	&	Wolinella succinogenes DSM 1740	\\
x1	&	Streptococcus thermophilus CNRZ1066	\\
x2	&	Streptococcus pyogenes M1 GAS	\\
x5	&	Bacillus cereus E33L	\\ \hline \hline
\end{tabular}
}
\caption{Abbreviations used for the genomes in Fig.~\ref{fig:crpVSpwl}.}
\label{GenName}
\end{table}

\end{document}